\documentstyle[12pt,aas2pp4,psfig]{article}

\newcommand{\al}{et al.}
\newcommand{\hi}{H{\sc i}}
\newcommand{\prim}{$^{\prime}$}
\newcommand{\prin}{$^{\prime\prime}$}
\newcommand{\aprox}{${\sim}$}
\newcommand{\Jy}{~Jy~km~s$^{-1}$}
\newcommand{\km}{~km~s$^{-1}$}
\newcommand{\degree}{$^{\circ}$}

\newcommand{\cg}{CGCG~160--}
\newcommand{\msolar}{M$_{\odot}$}
\newcommand{\rabell}{r$_{\mathrm A}$}

\newcommand{\mhdos}{M$_{\mathrm {H2}}$}

\newcommand{\defhi}{Def$_{\mathrm {HI}}$}

\begin{document}

\title{VLA HI imaging of the brightest\\ 
spiral galaxies in Coma}

\author{H. Bravo-Alfaro}
\affil{Observatoire de Paris DAEC, and UMR 8631, associ\'e au CNRS 
       et \`a l'Universit\'e Paris 7, 92195 Meudon Cedex, France.}
\affil{Departamento de Astronom\1a, Universidad de Guanajuato.
Apdo. Postal 144 Guanajuato 36000. M\'exico.}
\authoremail{hector@astro.ugto.mx}

\author{V.Cayatte}
\affil{Observatoire de Paris DAEC, and UMR 8631, associ\'e au CNRS 
       et \`a l'Universit\'e Paris 7, 92195 Meudon Cedex, France.}

\author{J. H. van Gorkom}
\affil{Department of Astronomy, Columbia University, New York, New York
10027.}

\and
 
\author{C. Balkowski}
\affil{Observatoire de Paris DAEC, and UMR 8631, associ\'e au CNRS 
       et \`a l'Universit\'e Paris 7, 92195 Meudon Cedex, France.}

\begin{abstract}
 
We have obtained 21~cm images of 19 spiral galaxies in the Coma
cluster, using the VLA in its C and D configurations.  The sample
selection was based on morphology, brightness, and optical diameters
of galaxies within one Abell radius (1.2\degree).  The
\hi~detected, yet deficient galaxies show a strong correlation in
their \hi~properties with projected distance from the cluster
center. The most strongly \hi~deficient (\defhi$>$0.4) galaxies are
located inside a radius of 30\prim ~(\aprox0.6 Mpc) from the center of
Coma, roughly the extent of the central X-ray emission. These central
galaxies show clear asymmetries in their \hi~distribution and/or
shifts between the optical and 21~cm positions.  Another 12 spirals
were not detected in \hi~with typical \hi~mass upper limits of 10$^8$
M$_\odot$.  Seven out of the twelve non detections are located in the
central region of Coma, roughly within 30\prim~from the center. 
The other non detections are to the east and southwest of the center.

Seven so called blue disk galaxies in Coma were observed in \hi~and
six were detected. These galaxies are relatively close to the central
region of Coma. The non detected one is the closest to the center.
The six detected blue galaxies are mildly \hi~deficient.  We did a
more sensitive search for \hi~from 11 of the 15 known post starburst
galaxies in Coma. None were detected with typical \hi~mass limits
between 3 and 7$\times$10$^7$ M$_\odot$.

    Our results present and enhance a picture already familiar for well
studied clusters. \hi~poor galaxies (deficient ones and non-detections) are
concentrated toward the center of the cluster. The \hi~morphology of the
central galaxies, with optical disks extending beyond the \hi~disks is unique
to cluster environments and strongly suggests an interaction with the IGM. A
new result in Coma is the clumpy distribution of gas deficiency. In the
cluster center the deficient galaxies are to the east while the
non-detections are to the west. In the outer parts the gas rich galaxies are
north of Coma, non-detected spirals are found in the NGC~4944 group to the
east and NGC~4839 group to the south west. This supports recent findings that
merging of groups is ongoing in the center of Coma, further out the NGC~4944
and NGC~4839 must have passed at least once through the core, while the
galaxies to the north have yet to fall in.

\end{abstract}


\section{Introduction.}

Two of the outstanding questions on cluster of galaxies are: what
causes the density morphology relation and how dynamically relaxed are
clusters. For the density morphology relation the question is: are all
galaxies formed by the same process and do they evolve differently in
different environments? Or does the environment play a dominant role
at the time of formation? On a larger scale the question is whether
clusters at low redshift are dynamically relaxed or whether they are
still accreting significant amounts of mass. The answer to this
question may contain a clue to the density of the universe (Thomas
\al~1998).  Observations of clusters at intermediate redshift suggest
that clusters evolve rapidly. Already at intermediate redshifts they
differ markedly from clusters at low redshifts.  Many clusters at
redshifts larger than z{\aprox}0.3 are known to contain a large
population of blue galaxies, indicative of enhanced star formation,
which is not found in similar clusters at z $<$ 0.1 (Butcher \& Oemler
1978, 1984; Dressler \& Gunn 1992, Poggianti \al~1999, Dressler
\al~1999).  Secondly these clusters often contain galaxies which have
recently undergone a starburst, but which are not currently forming
stars, displaying spectra with strong Balmer absorption lines but no
emission lines at all.  These galaxies are commonly referred to as
``post-starburst'' (PSB) galaxies.

We address both questions with \hi~imaging studies of nearby clusters,
which are directly comparable to more distant ones.  Detailed
\hi~imaging of the nearest cluster of galaxies, Virgo, has shown that
the neutral gas component gets affected dramatically by the hot
intergalactic medium (IGM) (van Gorkom \al~1984, Warmels 1988, and
Cayatte \al~1990, 1994).  In the center of Virgo the \hi~disks are
much smaller than the optical disks. The actual \hi~removal mechanisms
at work in the center of Virgo could be identified by a detailed
comparison of the \hi~surface density distribution of the cluster
galaxies with their counterparts in the field (Cayatte \al~1994).
Thus, the \hi~distribution of individual galaxies reflects the effect
of the cluster environment, while the location and velocities of
gas-rich versus gas-poor galaxies can help identify possible
substructure in the clusters (Valluri \al~1999). Gas can also be used to
probe the orbital history of the galaxies.  More recently other
clusters have been imaged in \hi~with the VLA and WSRT: Hydra (McMahon
1993), Ursa Majoris (Verheijen 1996), A~2670 (van Gorkom~1996), A~262
(Bravo--Alfaro \al~1997), and Hercules (Dickey 1997). Preliminary
results on Coma were obtained by Sullivan (1989) in the early days of
the VLA. In spite of poor sensitivity and interference problems in the
data, he observed disturbed \hi~disks for a few galaxies in the center.

In this paper we present a more complete and much more sensitive study
of the Coma cluster. We expect the environmental effects in Coma to be
stronger than in Virgo, because Coma is a much richer cluster and
because its central X--ray emission is almost six times more luminous
than in Virgo. Indeed, single dish observations have shown
that Coma is one of the most \hi~deficient clusters; extremely \hi~
deficient galaxies are seen out to 1.5~r$_{\mathrm A}$ (the Abell
radius, r$_{\mathrm A}$=1.7\prim/z, where z is the redshift), while in
other clusters strongly \hi~deficient galaxies are generally not seen
beyond 0.75~r$_{\mathrm A}$ (Bothun \al~1985; Giovanelli and Haynes
1985; Gavazzi 1987; 1989).  The Coma cluster is in fact the richest of
the nearby clusters (richness class 2, and z~=~0.023\footnote{We
assume a velocity of 7000~km~s$^{-1}$, and H$_{\circ}$~=~100
km~s$^{-1}$~Mpc$^{-1}$ throughout this paper.}), and represents the
closest equivalent to what is seen at intermediate redshifts, as it has blue
disk galaxies (Bothun \& Dressler 1986, and references therein) and
PSB galaxies (Caldwell, Rose, \& Dendy~1999, and references therein),
which are commonly seen in distant clusters. This makes Coma the perfect
link between nearby and more distant clusters.

In the present paper we discuss the global properties of Coma, on the
basis of our \hi~results, combined with recent observations on
different wavelengths and numerical work. A detailed analysis of
individual galaxies and more specific regions of Coma, will be the
subject of a forthcoming paper (Bravo--Alfaro \al~1999).  The plan of
this paper is as follows: Observations and data reduction are
described in Section 2. Observational results are given in Section
3. In Section 4 we discuss what the observations have taught us about
Coma at large. Conclusions are summarized in Section 5. A description
of \hi~properties of individual galaxies is given in Appendix
A.

\section{Observations}

We selected the 25 brightest spiral galaxies in Coma with
morphological type later than S0. All those galaxies have B~magnitude
smaller than 15.7~mag and optical diameters larger than
0.5\prim. Fig.~1 shows the location of the 12 fields observed
with the VLA,\footnote{The National Radio Astronomy Observatory is
operated by Associated Universities, Inc., under cooperative agreement
with the National Science Foundation.}  up to 1.2$^{\circ}$
(1~r$_{\mathrm A}$) from the center of Coma (we take the position of
NGC~4874 as the Coma center). Three fields (1, 2, and 9) are centered
around the core, covering most of the Coma velocity range. Field 8 was
pointed south of the elliptical NGC~4839, which is the dominant galaxy
of the SW group. Five other fields (3, 4, 6, 7, and 10) contain the
blue disk galaxies reported by Bothun \& Dressler (1986) and most of
the PSBs reported by Caldwell \al~(1993) and Caldwell \& Rose
(1997). The remaining three fields (5, 11, and 12) were centered in
the outer regions of the cluster, where no strong environmental
effects are expected, providing a comparison sample.

\placefigure{fig1}

The VLA survey of Coma was carried out in C and D array, during three
runs, in March 1995, April 1996, and March 1999. In March 1995, an
exploratory observation was done of fields 2, 4 and 5, in D array.
Twelve fields were observed in C array in April 1996.  Fields 1 and 10
were reobserved, March 1999, to get very sensitive observations of the
PSB galaxies in these fields. For the fields observed in both, C and D
configurations (fields 1, 2, 4, 5, and 10), the data were combined to
improve sensitivity. The combined data provide the higher spatial
resolution and a lower velocity resolution. In the D array
observations of 1995 online Hanning smoothing was used, after which
every other channel was discarded. This leaves a set of 31 independent
channels with a velocity spacing of 43 km/s.  In C array no online
Hanning smoothing was used, providing a set of 63 channels with a
velocity resolution of 21.7 km/s. Most of these observations were
afterwards Hanning smoothed to the same velocity resolution as the D
array data. The C array data were tapered to an angular resolution of
30 arcsec, and, in most cases, we obtained a resolution of 35 arcsec
for the combined C+D data.

Standard VLA calibration and imaging procedures were applied, using
the NRAO's astronomical image processing system (AIPS). Data cubes
were made using nearly pure natural weighting to obtain a higher
sensitivity.  A set of eight line--free channels on either side of the
band was used (four channels in the case of C+D combined data) to
define a mean continuum image. This image was then subtracted from the
channel maps forming an \hi~line emission cube. We used this cube to
search for 21~cm line emission and to determine the spatial position
and velocity range for each signal.  Finally, a new continuum image
was made using only line--free channels.  The channels containing line
emission were CLEANed (for more details on data reduction see
Bravo--Alfaro 1997).  The rms in the final cubes is typically 0.37
mJy/beam per channel. The observations have on average an \hi~mass
detection threshold of 10$^{8}$ M$_{\odot}$, corresponding to
6$\times\sigma\times$21.7km~sec$^{-1}$ (the channel width) and a
typical surface brightness sensitivity of 2 to 4$\times$10$^{19}$
cm$^{-2}$ (corresponding to 2.5~rms). Fields 1 and 10, were reobserved
in 1999 for 20 and 15 hours respectively; there our detection thresholds
are lower: the rms per channel is around 0.20~mJy/beam, and the
\hi~mass detection threshold is 2.4$\times$10$^{7}$ M$_{\odot}$ in the
center of the field.

The observational parameters are listed in Table~1, where
Column~1 indicates the observed field, Columns~2 and 3 give R.A. and
Dec. (1950) of each pointing, Column~4 gives the VLA configuration,
Column~5 the integration time, Column~6 the total bandwidth, Column~7
the heliocentric velocity of the central channel, Column~8 the channel
separation (in km~s$^{-1}$), Column~9 the rms noise per channel after
the continuum subtraction, and Column~10 the flux density per beam
area (in mJy/beam) equivalent to 1.0~K in the channel maps.

\placetable{tbl-1}

\section{Results}

We detected 19 galaxies in this survey; 17 spirals, one irregular, and
one interacting system. Table~2 lists those galaxies with
their observational parameters. Columns~1 and 2 give the galaxy
identification, Column~3 and 4 the R.A. and Dec. (1950) from the NED
data base, except (1) taken from the LEDA data base, Column~5 gives
the VLA configuration, Column~6 the synthesized beam size (in arcsec),
Column~7 gives the morphological type from Dressler 1980, except when
indicated: (1) from Huchra \al~1990, (2) from the LEDA database, and
(*) for uncertain classification. Column~8 gives the blue total
magnitude obtained from the LEDA database, except (1) taken from the
NED database; Column~9 gives the m$_{\mathrm {UV}}$-b color, from
Donas \al~1995.

In Table~3 we list the \hi~parameters derived from the
observations: Columns~1 and 2 are as in Table~2,  Column~3 gives
the observed field, Column~4 and Column~5 give the central
\hi~velocity and the velocity width, respectively. As the
signal-to-noise ratio is poor for some galaxies, we use as a
homogeneous criterium to obtain the central \hi~velocity 
the central channel displaying emission. The uncertainty in this value
is around half the velocity resolution, i.e. \aprox11\km.  The
velocity width is defined as given by the range of channels containing
H{\sc i} emission.  The error in this case is the velocity resolution,
around 22\km. Column~6 gives the \hi~total flux corrected for the
primary beam attenuation, with the corresponding error. Column~7 gives
the continuum intensity, Column~8 the total \hi~mass, Column~9 and 10
give the \hi~deficiency and the projected distance from the cluster
center respectively.

\placetable{tbl-2}
\placetable{tbl-3}

The main result of this paper is summarized in Fig.~2. It
shows a synthetic view of all the galaxies detected in \hi.  They are
placed at their correct location in the cluster and magnified by a
factor 7. Contours of the individual \hi~images (the first contour
corresponds to a column density of 3$\times$10$^{19}$~cm$^{-2}$) are
overlaid on optical images shown in grey-scale (DSS). The cross
indicates the center of the cluster (taken to be the position of
NGC~4874), and the central contours draw the X-ray emission in the
0.5--2 keV energy band, as observed with ROSAT (Vikhlinin, \al~1997).
This figure displays a wealth of information. As in Virgo, the first
thing to note is that the \hi~disks in the outer parts of the cluster
are much larger than the optical disks. But something different in
Coma, are the asymmetries and even displacements of the shrunken
\hi~disks seen in the center. All the shrunken HI disks are in
projection within the boundaries of the X--ray emission (see
Fig.~2), suggesting that an interaction with the IGM may be
at work. Near the cluster center, the distribution of \hi~detected
galaxies is very non uniform: most of the detections lie east
of the center, and there are almost no detections west of the
center. In the zone between the center of Coma and the SW group, only
one galaxy (Mrk 058), with very low gas content, was detected.

\placefigure{fig2}

Twelve galaxies from the original observed sample (morphological type later
than S0 and m$_{B}<$15.7) were not detected (Table~4). They are
within 20\prim~from the center of the observed fields and within the observed
velocity range.  Table~4 gives in Columns~1 and 2 the
identification, Columns~3 and 4 the R.A. and Dec. (1950), in Column~5 the
observed field, Column~6 gives the morphological type taken from Dressler
1980, except (1) taken from Huchra \al~1990, (2) from the NED database, and
(3) from the RC3 catalog; (*) means uncertain classification.  Column~7 gives
the optical velocity taken from the LEDA database, Column~8 the rms noise per
channel corrected for the primary beam response at the position of the
galaxy. The \hi~mass upper limit given in Column~9 corresponds to 6 times the
rms noise, multiplied by the channel width (typically 21.7~\km). Column~10
gives the lower limit to the \hi~deficiency.

\placetable{tbl-4}

In addition, we made a special effort to detect the so called abnormal
spectrum galaxies, reported in Coma by Caldwell \al~(1993) and
Caldwell and Rose (1997).  Some of them display ongoing star formation
activity (SB), and others a recent peak of star formation (PSB). Most
of these galaxies are found southwest of the cluster center (see
Sect.~3.4.2).  In Table~5 we give a list of all the peculiar
spectrum galaxies observed in this survey; none was detected in
\hi. Column 1 and 2 give the galaxy identification (the first
corresponds to the Dressler, 1980 catalog number), Column 3 gives the
observed field, Column 4 and 5 the (1950) R.A. and Dec.  Column 6 the
morphological type, and Column 7 the heliocentric velocity, both from
the NED database. Column 8 gives the rms noise per channel corrected
for the primary beam response at the position of the galaxy. Column 9
gives the \hi~mass upper limit for both, SB and PSB non detected
galaxies, typically between 3 and 7$\times$10$^{7}$~M$_{\odot}$. We
did not estimate the \hi~deficiency parameter, as most of these
galaxies are lenticular (morphological type S0).

\placetable{tbl-5}

\subsection{The HI Content}
\subsubsection{Comparison with previous works}

Five of the 19 detected galaxies are new \hi~detections.  Three of
them (FOCA~195, KUG~1255+275, and KUG~1258+287) are \hi~rich systems
with relatively low optical surface brightness, which have not been
observed previously in single dish \hi~surveys. NGC~4907 and \cg106,
were not detected in previous single--dish \hi~surveys probably
because of their low total \hi~flux (see Table~3). 
\cg106 was already detected with the VLA by Sullivan (1981). 
Four of the galaxies detected in the present survey
(NGC~4907, \cg106, Mrk~058, and IC~4040), were not detected in the most
recent survey carried out with the Arecibo telescope (Haynes \al~1997),
either because of low \hi~flux, or because several objects were found
inside the beam. 

In Fig.~3 we show a comparison between single dish and VLA
measurements.  There is in general a good agreement, showing that the
VLA has not missed any extended flux. Two discrepant cases are seen in
Fig.~3, NGC~4848 and IC~842. We obtained for the former,
systematicly a lower total \hi~flux than the values previously
reported (Chincarini et al. 1983, Giovanelli \& Haynes 1985, Gavazzi
1989 and Haynes \al~1997). We confirm the asymmetric distribution
found by Gavazzi (1989) but we probably miss some extended
\hi~emission. We obtained for IC~842 a slightly higher \hi~flux
(\aprox30\%) than previous single dish observations (Chincarini
\al~1983, Bothun \al~1985, Gavazzi 1989, and Haynes \al~1997).

We do not confirm two detections previously reported, Mrk~056 and
NGC~4944, which were only marginally detected by Gavazzi (1987). Our
\hi~flux upper limit for Mrk~056 (\cg064) is well below the flux
measured by Gavazzi. This galaxy is very close to the detected
\hi~rich galaxy KUG~1255+275, both spatially (4.6\prim) and in
velocity, thus this result may be due to confusion inside the Arecibo
beam. On the other hand, the detection of NGC~4944 is reported by
Gavazzi (1987) as a marginal one, and it was not confirmed in later
surveys (e.g. Gavazzi 1989).

\placefigure{fig3}

\subsubsection{ The HI deficiency}

To quantify the \hi~content of the spirals as compared to so called
field spirals we use the \hi~deficiency parameter (\defhi), following
the definition by Giovanelli \& Haynes (1985, and references therein),
where the deficiency parameter is the log of the ratio of the
average \hi~mass of isolated spirals of the same morphological type
and the observed \hi~mass. For our sample, we use the
morphological types from Dressler (1980), or from Huchra \al~(1990)
when the first is not available. We should remark that the
\hi~deficiencies in Coma are less well defined than for more nearby
clusters due to considerable uncertainty in the morphological
classification.

A diagram of the \hi~deficiency versus projected distance to the
cluster center is shown in Fig.~4, where all detected
galaxies are shown with black circles and lower limits for non detections
(only galaxies from Table~4 are included in this figure)
with triangles. The merging system NGC~4922 is not plotted in this figure
because of its peculiarity (see Appendix A).  Clearly the most
\hi~deficient galaxies are closer to the cluster core.  All strongly
\hi~deficient detected galaxies (Def$>$0.4) are inside a projected
radius of 30\prim~(0.4\rabell) from the center of Coma. For the non
detected galaxies the trend of deficiency with the projected distance
to the center is not as clear as it is for the detected ones, but 7 of
the non-detections are projected inside or very close to the X--ray
emission (UGC~8071, NGC~4851, IC~3943, NGC~4858, IC~3955,
KUG~1258+277, and \cg261), as shown in Fig.~5.

\placefigure{fig4}

\placefigure{fig5}

For the non detected galaxies we used 6 times the r.m.s. multiplied by
the velocity channel width to calculate both, the deficiency and the
\hi~mass upper limit shown in Table~4.  This value is
estimated on the basis of our detection threshold. Another method to
estimate the \hi~mass upper limit for non detected galaxies could
consider the expected velocity width for a big spiral, e.g. 300~\km,
rather than the observed channel width. In this fashion the total flux
upper limit would be obtained as r.m.s.$\times$2.5$\times$300~\km. But
as these galaxies are not in a typical environment we do not know at
all what the \hi~velocity width could be after a stripping
process. Using this different limit does not substantially change the
plot in Fig.~4. 

Although a possible correlation with distance to cluster center can be
considerably weakened by projection effects (Chamaraux \al~1980), the
fact that the most deficient galaxies are almost all situated within
the central and southwestern X--ray emission suggests that an
interaction with the IGM is at work. More convincing evidence for this
can be obtained by looking at the relative sizes of the \hi~disks. No
other mechanism would cause the \hi~disks to be smaller than the
optical ones. We will address these issues in a following paper.

\subsection{The central region}

Fig.~5 shows the position of the detections with crosses, and
the 12 non detected galaxies with triangles. We consider here as non
detections those galaxies brighter than a magnitude of 15.7, with
morphological type later than S0 (Table~4).  We also show in
Fig.~5 the contours of the X--ray emission (ROSAT) centered
on Coma, as reported by Vikhlinin \al~(1997). The central region of
Coma contains most of the strongly \hi~deficient galaxies in the
cluster, including the very bright early spirals NGC~4911, NGC~4921,
and NGC~4907, all of them very \hi~deficient. The giant spiral
NGC~4911, detected in X-ray, is thought to be the dominant galaxy of a
group crossing the main Coma body.  NGC~4921 (classified as ``anemic''
by van den Bergh, 1976) and NGC~4907 show very perturbed \hi~maps;
they have large velocities relative to the mean cluster velocity (1179
and 1521\km~respectively). Four other galaxies were detected inside a
radius of 30{\prim}~(\aprox 0.60 Mpc) which is roughly the area
defined by the X--ray emission shown in Fig.~5. With no
exception these galaxies show deeply perturbed \hi~distributions, some
of them smaller than the corresponding optical disks (see Fig.~2).

\subsection{The SW of Coma}

The presence of a group in the SW of Coma, associated with the giant
elliptical galaxy NGC~4839, has been well established through optical
and X--ray studies (e.g. Briel \al~1992, White \al~1993, Colles \&
Dunn 1996, Biviano \al~1996). The group sits 40\prim~SW of the cluster
core, and coincides with the second most intense X--ray peak. This
group represents 6\% of the total X--ray emission, and contains about
10\% of the mass of the cluster. The question whether this group is
falling into the center of Coma (Colles \& Dunn 1996), or has already
passed following a straight path (Burns \al~1994), is still a matter
of debate. Recently, Caldwell \& Rose (1997) concluded that the SW
group has already passed through the main body of Coma, based on the
velocity structure of the group and the presence of PSB galaxies in
that zone. Our field number 8, which includes NGC~4839, and field
number 10 in the zone between this galaxy and the cluster center, try
to clarify this controversy.

Of the 18 galaxies surrounding NGC~4839 and belonging to the group
(Biviano 1998), four were observed and none were detected. This is
perhaps not too surprising since all of them are classified as S0. We
did however detect three late type galaxies just south of the NGC~4839
group (IC~3913, KUG~1255+275 and Mrk~057), just outside the X--ray
emission (Figs.~\ref{fig2} and~\ref{fig5}). Their radial velocities
are close to the systemic velocity of the group (7339${\pm}$329~\km,
Colless \& Dunn 1996), thus they are likely to be group members. These
galaxies have undisturbed morphologies and a normal \hi~content, which
makes it extremely unlikely that they have crossed the cluster core.

Our \hi~results are inconclusive. The non detections in the close
vicinity of NGC~4839, and the presence of several SB and PSB galaxies
in that zone with a very low \hi~content (see Sect.~3.4.2 and
Table~5), support the hypothesis that the group has gone
through. The 3 gas rich galaxies pose a problem. If they are members
of the group, the group cannot have passed through the main
cluster. Alternatively they may be new members of the group, only
recently accreted after the group has gone through the cluster center.

\subsection{HI content of SB and PSB galaxies in Coma.}
\subsubsection{The blue disk galaxies}

As mentioned before Coma contains blue disk and post--starburst
galaxies, similar to those seen at intermediate redshifts. Bothun \&
Dressler (1986) obtained spectroscopy for seven blue galaxies in Coma,
and showed that they had not only experienced a star burst, but were
still forming stars at a high rate. Considering various mechanisms to
produce the starburst phenomena these authors conclude with a mild
preference for ram pressure induced star formation. The main reason
for rejecting galaxy--galaxy induced star formation is that the blue
disks do not appear to be interacting (except the non detected
NGC~4858, possibly interacting with NGC~4860 and member of an {\it
aggregate}, Conselice and Gallagher 1998).  All seven blue disk
galaxies were observed in this survey and six of them were detected:
NGC~4848, Mrk~058, \cg086, IC~4040, \cg098 and NGC~4926-A. The first
four are projected inside the X--ray emission (Fig.~6) and
are \hi~deficient (\defhi $>$ 0.5). They display very perturbed
\hi~distributions. The \hi~deficiencies and gas distribution strongly
suggest that these galaxies are being ram pressure swept by the IGM.

The locations of the \hi~detected blue disk galaxies are indicated with
crosses in Fig.~6. Most of them are in an annular region between
20\prim~and 30\prim~from the cluster center, which coincides with the
outermost contour of the X--ray emission (corresponding to a gas density of
\aprox3$\times$10$^{-4}$~cm$^{-3}$).  This is similar to what has been seen
in Butcher--Oemler clusters, where blue galaxies seem preferentially located
in an annular region outside the cluster core (e.g. Butcher and Oemler 1984;
Mellier \al~1988). Dressler \& Gunn (1990) reported that star forming
galaxies seem to avoid the cluster core, appearing first at a radius of
\aprox0.5 Mpc (0.5 Mpc=25\prim~in Coma). Furthermore, the UV survey of Coma
by Donas \al~(1995) revealed that 38\% of the UV flux is produced in a ring
lying between 20\prim~and 30\prim~of the cluster center. This suggests that
this effect must be related to the global properties of the cluster, perhaps
IGM induced star formation can best explain the shell of blue starburst seen
in Coma and similarly in high redshift clusters (Oemler 1992).

\placefigure{fig6}

We also show in Fig.~6, with triangles, the non-detected blue
disk galaxy (NGC~4858) and three starburst like objects reported by
Caldwell \al~(1993), which were also not detected in this survey
(Table~5). The blue disk is projected well inside the X--ray
emission and star formation followed by ram pressure stripping may have
exhausted their \hi~reservoir. The other three galaxies show abnormal
spectra with strong emission lines (Caldwell \al~1993). These objects
are also near the edge of X--ray emission, in the SW direction. Their
non detection suggests that in this zone, between the center and the
SW group, the gas is more easily removed. Note that starbursting
galaxies in other directions (e.g. NGC~4848, IC~4040, and \cg086), at
similar distances from the center, are detected in \hi.

\subsubsection{The post starburst galaxies}

Caldwell \al~(1993) and Caldwell \& Rose (1997) reported a total of 22
abnormal spectrum galaxies in Coma. Five of them display Balmer
absorption and emission lines (see previous paragraph), two AGN's, and
15 galaxies show enhanced Balmer absorption lines but no emission,
similar to PSBs observed in z\aprox0.3 clusters. Most of the PSBs in
Coma are early type objects (mostly S0's). The spectra indicate that a
burst of star formation was recently truncated (\aprox1 Gyr
ago). Among the 15 PSBs 8 are in the SW region, the remaining are near
the center and in the NE (Caldwell \al, 1993). We observed
11 of those galaxies, and detected none (see Table~5). The
position of those galaxies are shown in Fig.~6 with dotted
circles. The X-ray contours (Vikhlinin \al~1997) are also drawn.

The NGC~4839 group has an average velocity close to the peak in the
PSBs velocity distribution (Biviano \al~1996). This suggests that the
PSBs could be part of the group and they have been stripped of their
gas when they passed through the Coma core. The few PSBs in the
center and NE of Coma could be old members of the SW group, previously
pulled out from the group when it passed through the cluster center
(Caldwell \al~1993).

Our results suggest an evolutionary sequence, where galaxies first
become blue because of IGM induced star formation. Star formation and
further ram pressure stripping makes them \hi~deficient. The displaced
and shrunken \hi~disks of the blue disk galaxies indicate that some
interaction with the IGM is indeed going on. The next step in this
evolutionary sequence is the PSB phase, where galaxies have lost most
of the \hi~gas, and as consequence, star formation stops on a
relatively short time scale. In this picture all the PSBs would be in
an advanced stage of gas stripping.

\subsubsection {Correlation between Colors and HI--Deficiency}

Although gas is needed for star formation it is not obvious that only
the atomic gas content is related to star formation activity (Donas
\al~1990). Molecular clouds are denser and more centrally located in
the galaxies, making it less likely that this component will be
affected by the cluster environment. For example Kenney \& Young 1989
(and references therein) showed in Virgo, and Casoli et al. (1991) in
the Coma supercluster, that there is no evidence for a lower molecular
content than in the field, even for the \hi~deficient galaxies. For
instance, the starburst NGC~4858, which is not detected in \hi~in the
present survey (\defhi~$>$0.93), shows normal CO emission. The same is
true for the detected, yet \hi~deficient galaxies such as NGC~4848,
NGC~4911, and NGC~4921. On the basis of new CO data for galaxies in
Coma, Boselli \al~1997 conclude that no correlation is observed
between H$_2$ content and the \hi~deficiency parameter, and that
molecular gas is not perturbed by the interaction with the IGM. More
recent CO observations for galaxies in Coma, by Lavezzi \al~(1999),
give additional support to that conclusion.

While CO and \hi~do not correlate, we show in Fig.~7 that in Coma
star formation does appear to correlate with \hi~deficiency. In that figure
we display the \hi~deficiency versus the mUV--b color, a very good tracer of
star formation.  Sixteen of the 19 galaxies in our sample have been observed
by Donas \al~(1995) with the imaging UV telescope FOCA (see mUV--b values in
column 9 of Table~2). A clear correlation appears in this plot,
where the linear fit is shown (the correlation coefficient is 0.61). Points
in the bottom right corner of Fig.~7 correspond to the three central
early spirals (NGC~4907, NGC~4911, and NGC~4921). This trend, where galaxies
with red mUV--b colors are the most \hi~deficient, has also been found in
Virgo by Giovanelli and Haynes (1983). Thus apparently colors become redder
if the interaction with the IGM is effectively stripping the \hi~gas.  As a
comparison, we show in Fig.~8 a plot mUV-b vs. the \mhdos~surface
density (drawn by the ratio between \mhdos~and the optical disk area) for
seven of the most \hi~deficient galaxies in Coma. This figure shows a weak
correlation between the mUV-b color and the molecular content, a high
dispersion is seen (the correlation coefficient is only 0.47).

It is worth mentioning that the clear correlation shown in
Fig.~7 is not due to a color--morphology trend, as
\hi~deficiency occurs both in early and late spirals. We tested for a
possible correlation between the \hi~deficiency and the morphological
type, and the result is that no trend is found (see Column~7 in
Table~2 and Column~9 in Table~3). These results suggest that in Coma
as in Virgo (Giovanelli and Haynes 1983) stripping has affected the
star formation history of the galaxies.

\placefigure{fig7}

\placefigure{fig8}

\subsection{The periphery of Coma; the East and Northern regions.}

Three fields were observed in the Coma periphery. The eastern field
(Field 11) was centered on the early type spiral NGC~4944, for which
we do not confirm Gavazzi's (1987) detection. Quite surprisingly two
other non detections are in this region. Biviano \al~(1996) propose
that NGC~4944 is the center of one of the groups in the Coma
structure. The general \hi~deficiency of this group suggests that it
already passed through the cluster center.

In the northern region, the \hi~data confirm the presence of two
other groups.  The first of them, observed in Field 6, is possibly
associated with the blue disk galaxy \cg098. We detected 3 \hi~rich
galaxies in that zone (\cg098, FOCA~195, and KUG~1258+287).  They are
separated from each other by distances around 13\prim ~(\aprox 260
kpc), and their velocities range from 8426\km~to 8884\km. Ours is the
first redshift determination of FOCA~195.  We also see \hi~emission,
still to be confirmed, 1.4\prim~W of KUG~1258+287, which does not seem
to have an optical counterpart on the DSS.  The total \hi~mass is
2.2$\times$10$^8$~\msolar, a typical value for a dwarf galaxy.

The last field (Field 5) was pointed \aprox1.2\degree~NE of the Coma
center. In this field we detected two \hi~rich galaxies, IC~842, and IC~4088,
and the interacting system NGC~4922, located 84\prim~away from the cluster
center. The pair NGC~4922 is the brightest IR object in Coma (Mirabel \&
Sanders 1988), composed of a brighter spiral component in the north, and a
fainter lenticular in the south.  In addition, the spiral has a central radio
continuum source, 5C4.130, with a flux of 60 mJy (Wilson 1970).  We detect
weak \hi~emission and confirm the strong absorption feature reported by
Gavazzi (1987). We confirm that the \hi~absorption and emission features are
centered on the northern spiral component.

Three \hi~rich dwarf systems were detected in field 5, associated with
the spiral IC~4088. Their \hi~maps look quite regular.  The dwarf
galaxies are previously uncatalogued in \hi, except [GMP83]~1866,
which was reported by Sullivan (1989). This object shows a total
\hi~mass of \aprox2.1$\times$10$^{9}$~\msolar, and a velocity width of
173~\km. The two other dwarfs have \hi~masses in the range
2.2--6.1$\times$10$^{8}$~\msolar, and velocity widths of 43~\km~and
130~\km.  The position of these galaxies and the absence of \hi~rich
dwarfs close to the center of Coma, suggests that these systems tend
to be present only in the outskirts of the cluster, associated with
luminous galaxies. This is supported by the fact that we devoted only
two fields to observe regions far from the X-ray source of Coma, and
it is there where \hi~rich dwarfs appear, while in the remaining more
central ten fields (a 5 times larger observed volume), we detected
none of these objects.

\section{Discussion}

The goal of our \hi~imaging studies of clusters of galaxies is twofold:
to investigate the interaction between individual galaxies and the cluster
environment and to probe the dynamical state of the cluster. Physical
mechanisms affecting individual galaxies will be discussed in a second paper
(Bravo-Alfaro \al~1999), here we place the observational results in the 
larger context of the cluster evolution.

Several recent papers have discussed the dynamical state of Coma based on the
optical mass distribution (Girardi \al~1994, Biviano \al~1996, Colles and
Dunn 1996), the surface brightness distribution of the X--ray emission
(Vikhlinin \al~1997 and references therein) and the temperature structure of
the X ray gas (Donnelly \al~1999 and references therein). The cluster
\hi~distribution shown in Figures~2, 5 and 6, strengthens some of the arguments
used by those authors. These figures show gas rich galaxies associated with
the \cg098 and IC~4088 in the north, a lack of \hi~in the groups associated
with NGC~4944 in the east, NGC~4911 in the center and NGC~4839 in the
south. The \hi~deficient galaxies are east of the center, while the non
detection are mostly west of the center. This result is consistent with the
recent analysis by Donnelly \al~(1999), who find that there is cooler gas
just south and south east of the center, while there is a hot spot in the
X--ray gas just north of the center.  These authors argue that most likely
the NGC~4874 group is a recent arrival in Coma, leaving behind a cool trail
of gas, the hot spot would then be the bow shock generated as the group moves
through the IGM. Interestingly numerical simulations show (Roettiger
\al~1996) that in minor mergers the infalling object will first develop a
protective bow shock, which will shelter the galaxies from being ram pressure
stripped. Close to core passage the infalling object's ISM encounters a
rapidly increasing density, leading to a burst of ram pressure stripping.
This may explain that we do detect galaxies east and south east of the
center, though they are \hi~deficient, while our non detections are mostly
west of the center and those galaxies have probably already passed the core.
Note that several of the small \hi~disks in Coma are very asymmetric,
indicative of current stripping, since any asymmetry would be smeared out in
roughly one rotation period (\aprox10$^8$ years).

A question that remains is what triggered and stopped the starburst in the
blue disk galaxies and PSBs respectively.  Several mechanisms have been
suggested, galaxy--galaxy interactions (Combes \al~1988), galaxy harassment
(Moore \al~1996, 1999), potential of a group--cluster interaction (Bekki
1999), and ISM--IGM interactions (Poggianti \al~1999 and references
therein). Least likely of those is the galaxy--galaxy interaction
scenario. The blue disks are located close to the center, in the outskirts of
the X ray source and show no optical signs of being disturbed.  Our
\hi~observations showed that it is the galaxies in the outer parts of the
cluster that have gas rich companions, while most of the center galaxies
appear more isolated. We investigated this further by counting the number of
galaxies inside a 5~arcmin circle around each of the galaxies in Tables~2 and
3. Five arcmin corresponds to 100 kpc at the distance of Coma, encounters at
larger distances are unlikely to be effective (Moore \al~1999 suggest that an
encounter with an impact parameter of 60 kpc is typical in a rich cluster of
galaxies). The number of neighbours was taken from NED, which unfortunately
has rather incomplete velocity information.  In Figure~9 we plot mUV-b color
versus number of close neighbours.  Clearly no correlation is seen in this
plot, and blue galaxies do not show an excess of close neighbours (if the
weak trend in the opposite direction was real, it would mean that red objects
have a slightly larger number of neighbours).

The location of the blue disk galaxies and the observation that the UV
radiation peaks in an annulus 20~arcmin from center (Donas \al~1995),
something which has also been found in more distant clusters (Butcher and
Oemler 1984, Mellier \al~1988, Dressler \& Gunn 1990, Dressler \al~1999),
makes it more likely that the starbursts are caused by gas--gas interaction
or an interaction with the global cluster potential.  The fact that the blue
disk galaxies are \hi~deficient and have lost their gas in the outer parts,
makes an ISM--IGM interaction slightly more plausible than the interaction
with a tidal field (Bekki 1999). Tidal interactions tend to move gas toward
the center and outward (Barnes,1998), thus even if it is the tidal
interaction that causes the burst, the gas in the outer parts must still have
been removed some other way. Numerical work by Stevens \al~(1999) shows that
interaction between galaxies and the IGM may produce mass loss and features
like bow--shocks and stripped tails (observed in X--ray), that should be
present in cluster with different degrees of richness.  These effects are
well known to prelude star formation activity.  Observationally, most of the
detected blue disks in the present survey are projected inside or at the edge
of the X-ray emission and they seem to have lost their \hi~in the outer
parts, while the non detected blue disk (NGC~4858) sits even deeper within
the X--ray source. It is especially the loss of \hi~in the outer parts of
galaxies that is suggestive. Of the PSB galaxies none is detected in \hi, and
most of them are located within the X-ray source.  Thus it seems plausible
that in Coma at least the starburst gets triggered by an interaction with the
IGM and the starburst is likely to stop because most of the remaining gas
gets swept out of the galaxies.

\section{Summary}

We carried out a 21~cm survey of the Coma cluster with the VLA.  High
resolution images were obtained for 19 spiral galaxies inside
1\rabell~(1.2\degree). Five of the galaxies in the present
survey are new detections in 21~cm. 

\noindent
-- We find gas rich and gas poor groups of galaxies in Coma that can
be isolated in space and velocity. The groups north of the cluster are gas
rich while groups associated with NGC~4944 to the east and NGC~4939 to
the southwest are \hi~poor. Even in the cluster center asymmetries
occur, the galaxies west of the center are very HI deficient with
small \hi~disks, while the galaxies east of the center are not detected
at all in \hi.  We suggest that the most \hi~poor groups have gone
through the center, while the deficient galaxies east of the center
are currently falling and being stripped.

\noindent
-- As expected, the environmental effects on the \hi~properties of spirals
in Coma are stronger than those previously reported in Virgo, where the IGM is
less dense than in Coma.

\noindent
-- We confirm the tendency of \hi~deficient galaxies to be closer to the
cluster core: seven of the detected galaxies are very \hi~deficient
(\defhi$>$0.4), and they lie inside a radius of 30\prim~from the cluster
center. This zone roughly coincides with the X--ray emission. With no
exception these galaxies show very perturbed \hi~distributions. For most of
them ram pressure by the dense IGM is a likely explanation of their abnormal 
\hi~properties, but also other mechanisms, such as viscous stripping and
conversion of \hi~to molecular gas, are probably present.

\noindent
-- Six blue disk galaxies are detected in Coma in \hi. Four of them are
\hi~deficient and they are projected inside an
annular zone between 20\prim~(0.4 Mpc) and 30\prim~(0.6 Mpc) from the cluster
center. This has been observed in high redshift clusters 
where star forming galaxies seem to avoid the cluster core, appearing first
at a radius of \aprox0.5 Mpc. This annulus roughly coincides in Coma with
the outer region of the X--ray emission as well as the peak of UV
emission. This confirms a close correlation between the location
of star forming galaxies and global properties of the cluster.

\noindent
-- Eleven PSBs were observed in this survey but none was detected;
\hi~mass upper limit as low as 3$\times$10$^{7}$~\msolar~are
found. Most of these galaxies are projected onto the hot IGM medium
outlined by the X-ray emission.

\noindent
-- We find a correlation between \hi~deficiency and UV-b color, in the
sense that the most deficient galaxies are reddest. This suggests 
that the loss of atomic gas reduces the star formation rate, despite
the fact that the molecular gas seems hardly affected.

\acknowledgments

\noindent
We thank the NRAO for generous allocation of observing time and the
VLA staff for help with the observations. HBA thanks the CONACYT of
Mexico, for its support through a Ph.D. grant, and the Mexico--USA
Foundation for Science, for its support through a summer
grant. HBA also thanks the DAEC of the {\it Observatoire de Paris},
the Astronomy Department of Columbia University, and the AOC of the
NRAO, for its support and hospitality during his visits. JHvG was
supported by an NSF grant to Columbia University. JHvG and HBA thank
the organizers of the 1998 Guillermo Haro Workshop on the Formation
and Evolution of Galaxies at INAOE, where part of this work was
done. We have made use of the Lyon-Meudon Extragalactic Database
(LEDA) supplied by the LEDA team at the CRAL-Observatoire de Lyon
(France). We used NED, the NASA/IPAC extragalactic database, operated
for NASA by the Jet Propulsion Laboratory at Caltech. We used the
Digital Sky Survey, produced at the Space Telescope Science Institute.

\newpage

\appendix
 \section{Description of \hi~properties}

\begin{itemize}

\item {\bf IC~3913} This late type spiral is one of the three galaxies
detected in the SW of Coma, lying at 17\prim~(0.34~Mpc) SW of
NGC~4839. IC~3913 shows a normal \hi~content (\defhi=0.04) and a
regular \hi~distribution). Its most external \hi~contour displays a
clear extension to the east, what accounts for the asymmetry observed
by Gavazzi (1989) from single dish observations.

\item {\bf NGC~4848} This blue disk Scd galaxy, projected on the NE
edge of the X-ray source, shows one of the most perturbed gas
distribution of this survey, and a high \hi~deficiency
(\defhi=1.14). However, this value should be taken with care, as
Fig.~3 shows that our \hi~flux value is systematicly lower
than in previous single dish surveys, what suggests we lost some flux
in the central region.  Nevertheless, the asymmetric \hi~distribution
most be real, with most of the neutral hydrogen in the north as
reported by Gavazzi (1989).

\item {\bf CGCG~160--058} This late type spiral galaxy, in the NE of
Coma, shows an \hi~distribution extending slightly further than the
optical disk. We obtained a normal \hi~content (\defhi=0.12), contrary
to Sullivan (1989), who obtained a smaller \hi~extent than the optical
disk and a slight \hi~deficiency (\defhi=0.4).

\item {\bf KUG~1255+275} This irregular galaxy, in the SW of Coma, is
detected for the first time in 21~cm.  It is an \hi~rich galaxy
(\defhi=--0.17), with rather regular gas distribution.

\item {\bf Mrk 057} This galaxy is located SE from NGC~4839. It is a
very \hi~rich galaxy (\defhi=--0.44), with an \hi~distribution
displaying an extension to the north.

\item {\bf Mrk 058} This blue disk galaxy, in the region between the
SW group and the center of Coma, shows a very asymmetric gas
distribution.  Most of the \hi~gas is placed in the west side, while
the east appears depleted. It has a high relative velocity to the main
cluster (\aprox1500~\km).

\item {\bf CGCG~160--076} This \hi~rich galaxy, with Def=-0.65, is
placed at 40\prim~(0.8~Mpc) in the north of Coma.  It was previously
tentatively imaged in 21~cm by Sullivan (1989), who reported an offset
between the optical and \hi~positions. We do not confirm this result.

\item {\bf CGCG~160--086} This blue disk galaxy is projected onto the
SE edge of the X--ray emission. The detection of
this galaxy is less reliable than the others because the data cube was
contaminated by interference close to the position of the galaxy.

\item {\bf IC 4040} This blue disk galaxy is, among our detections, the 
closest to the cluster center. It is \hi~deficient (\defhi=0.61), and
its neutral gas distribution is asymmetric, with most of the \hi~in
the SE, while the NW appears depleted.

\item {\bf IC~842} This late type spiral is in the NE of Coma,
\aprox1$^{\circ}$ from the cluster center. Its \hi~distribution is
regular and slightly larger than the optical disk.  We obtained a
rather normal \hi~content, \defhi=0.31, compared with previous
observations (see Sect.~3.1).

\item {\bf KUG~1258+287 \& FOCA~195} These two galaxies, very close to
each other, are detected for the first time in 21~cm. They are placed
in the north, at \aprox36\prim~(0.72~Mpc) from the cluster center. Our
velocity estimation for FOCA~195 is the first ever reported. Regular
gas distributions, and normal \hi~contents for both galaxies were
found.  We detected an independent \hi~emission, still to be
confirmed, 1.4\prim~W of KUG~1258+287, with no optical counterpart
seen in the DSS. We estimate a total \hi~mass of
2.2$\times$10$^{8}$~\msolar~for this detection.

\item {\bf NGC~4907} This bright Sb spiral, projected at only
19\prim~(\aprox0.38~Mpc) NE of the Coma core, is another newly
\hi~detected galaxy. The \hi~map shows a
\hi~disk clearly smaller than the optical one, what is expected for
such an \hi~deficient galaxy: \defhi=0.99. It appears a striking
feature extending through 24\prin~(\aprox 8~kpc) to the NW of the
galaxy, but as this emission appears in only one independent channel
it should be confirmed by future observations.

\item {\bf NGC~4911} This is one of the two giant spirals in Coma,
projected at 20\prim~(\aprox 0.4~Mpc) SE of the core. It is
\hi~deficient (\defhi=0.58), and presents an offset of 5--10 arcsec
between \hi~and optical positions. A striking emission is detected SW
of this galaxy (not shown in Fig.~2), what suggests a close
interaction with its neighbour DRCG~27-62. We will discuss this
result in a future paper.

\item {\bf NGC~4922/KPG~363--B} This interacting pair lies at the
northern periphery of Coma. An absorption and an emission \hi~features
are present, both coinciding with the northern spiral galaxy. Gavazzi
(1987) reported two emission features in both sides of the absorption
line, but we detect the emission only in the lower velocity side
(6892~\km). This emission is seriously contaminated by the absorption,
and we obtained a total flux of 0.12\Jy, only half of the value
reported by Gavazzi (1989).

\item {\bf CGCG~160--098} This is the \hi~richest galaxy among the
blue disks reported by Bothun \& Dressler (1986) in Coma. It is the
only blue disk galaxy located outside the X--ray emission, 50\prim~NE
from the cluster center. Its \hi~map extends well beyond the optical
disk and it shows a normal \hi~content (\defhi=0.16).

\item {\bf NGC~4921} This is the other giant spiral in Coma, also 
projected close to the cluster center, at 25\prim~SE (\aprox0.5
Mpc). We obtain a very high \hi~deficiency: \defhi=1.11.
It shows a very perturbed gas distribution, which is
clearly less extended than the optical disk. Most of the \hi~emission
is distributed along the SE spiral arm, while the NW appears
depleted. The \hi~centroid exhibits a slight offset, 10\prin~E of the
optical center.

\item {\bf IC~4088} Observed in the far northern field, this galaxy
shows a normal \hi~content (Def=-0.01), and a regular
\hi~distribution. Three dwarf systems, clearly detected in \hi,
encircle this galaxy; optical DSS counterparts are associated with the
dwarfs.

\item {\bf NGC~4926--A} Blue disk galaxy projected near the SE edge of the
X--ray source. It shows a normal \hi~content (Def=0.16) and a similar
extension for optical and \hi~disks.  Our 21~cm velocity (V=6876~\km) is
significantly lower than the optical velocity reported (7100~\km, Amram
\al~1992).

\end{itemize}

\newpage

\figcaption[bravo.fig01.eps]{Plot of the 12 observed fields in Coma. The size
      		of the circles indicates the effective field of view of each
      		pointing. The segmented circle outlines the 1~Abell radius of
      		Coma (1.2 degrees).  Crosses indicate detected galaxies and
      		the asterisk indicates the cluster center, coincident with
      		the giant galaxy NGC~4874.\label{fig1}}

\figcaption[bravo.fig02.eps]{Composite plot of individual \hi~maps of
                   spiral galaxies of Coma observed with the
                   VLA. Galaxies are shown at their proper position
                   (except those in the rectangle, where the position
                   of \cg102 is indicated with an ``x'') and they are
                   magnified by a factor of 7. The \hi~maps are
                   overlaid on DSS optical images. The first contour
		   is the same for all galaxies,
                   corresponding to $3 \times 10^{19}$ cm$^{-2}$.  Their
                   identification and central velocities (in \km) are
                   indicated. The large scale contours sketch the
                   X--ray emission as observed by Vikhlinin
                   \al~1997. The cross indicates the cluster center,
                   coincident with the elliptical
                   NGC~4874. \label{fig2}}

\figcaption[bravo.fig3.eps]{Comparison of total \hi~flux obtained in
                   this work with previous 21~cm observations. The
                   dotted line indicates the 45$^{\circ}$ fit
                   position. Solid lines join different estimations
                   done for the same object.  Reference keys are:
                   Gavazzi 1987 (G87), Gavazzi 1989 (G89), Giovanelli
                   and Haynes 1985 (GH85), Bothun \al~1985 (B85),
                   Chincarini \al~1983 (C83), and Haynes \al~1997
                   (H97) \label{fig3}}

\figcaption[bravo.fig4.eps]{The distribution of the \hi~deficiency parameter
                  as a function of the projected radius from the center of
                  Coma.  Black circles correspond to detected galaxies, and
                  triangles to lower limits for non detected. \label{fig4}} 

\figcaption[bravo.fig5.eps]{The position of the detected (crosses) and
                 non--detected (triangles) galaxies in the present survey,
                 superposed on the ROSAT X--ray emission (Vikhlinin
                 \al~1997). \label{fig5}}

\figcaption[bravo.fig6.eps]{The position of the detected (crosses) and
                 non--detected (triangles) starburst galaxies, as well as
                 the non--detected (dotted points) PSBs. They
                 are superposed on the ROSAT X--ray emission (Vikhlinin
                 \al~1997). \label{fig6}}

\figcaption[bravo.fig7.eps]{The plot mUV-b color versus the
                  \hi~deficiency parameter. The linear fit is
                  indicated. A clear correlation is seen in spite of
                  the high dispersion (the correlation coefficient is
                  0.61).  \label{fig7}}

\figcaption[bravo.fig8.eps]{The plot mUV-b color versus the molecular
                  gas content, given by the ratio M$_{H2}$/optical
                  disk area (pc$^2$). A weak correlation is
                  found (correlation coefficient of 0.47).
                  \label{fig8}}

\figcaption[bravo.fig9.eps]{Plot of mUV-b color versus the number of
                  neighboring galaxies inside a 5\prim~radius (from
                  NED). Blue galaxies do not have a larger number of
                  close neighbours than red ones, making it unlikely
                  that major galaxy--galaxy interactions have triggered
                  the star burst. \label{fig9}}

\newpage

\begin{deluxetable}{ccccrccccc}
\footnotesize
\tablenum{1}
\tablecaption{Observational parameters \label{tbl-1}}
\tablewidth{0pt}
\tablehead{
  \colhead{(1)}  & \colhead{(2)}      & \colhead{(3)}        
& \colhead{(4)}  & \colhead{(5)}      & \colhead{(6)}
& \colhead{(7)}  & \colhead{(8)}      & \colhead{(9)}     & \colhead{(10)} \nl 
  \colhead{Field}                         & \colhead{${\alpha}_{\small 1950}$} 
& \colhead{${\delta}_{\small 1950}$}      & \colhead{Config}  
& \colhead{t$_{\mathrm {int}}$}           & \colhead{Bandwidth}    
& \colhead{Central vel}                   & \colhead{${\Delta}$v}   
& \colhead{rms}                           & \colhead{mJy--K conv}   \nl   
  \colhead{}              & \colhead{h m s}                  
& \colhead{$^{\circ}$ $^{\prime}$ $^{\prime\prime}$} 
& \colhead{VLA}           & \colhead{hrs}                
& \colhead{MHz}           & \colhead{km~s$^{-1}$}    
& \colhead{km~s$^{-1}$}   & \colhead{mJy/beam} 
& \colhead{mJy/beam}       
}
\startdata
1 & 12 58 44.8                   & 28 18 05.0  & C
& 4.0                  & 6.25           & 7800        
& 21.5          & 0.40     & 1.29           \nl 
  &                              &             & D
& 20.0                 & 6.25           & 7650
& 21.5          & 0.20     & 5.47           \nl
2 & 12 58 30.0                   & 28 18 00.0  & C
& 4.0                  & 6.25           & 5500        
& 21.5          & 0.39     & 1.43           \nl 
  &                        &                   & D
& 2.0                  & 6.25           & 5500        
& 43.3          & 0.32     & 4.81           \nl
3 & 12 55 24.4                   & 28 43 10.5  & C
& 4.0                  & 6.25           & 7300        
& 21.5          & 0.42     & 1.32           \nl 
4 & 12 59 05.0                   & 28 00 00.0  & C
& 4.0                  & 6.25           & 7200        
& 21.5          & 0.40     & 1.42           \nl 
  &                              &             & D
& 2.0                  & 6.25           & 7200        
& 43.3          & 0.35     & 4.85           \nl
5 & 12 59 00.0                   & 29 23 58.0  & C
& 4.0                  & 6.25           & 7000  
& 21.5          & 0.38     & 1.42           \nl 
  &                              &             & D
& 2.0                  & 6.25           & 7000  
& 43.3          & 0.29     & 4.84           \nl 
6 & 12 59 00.8                 & 28 47 28.0    & C
& 4.0                  & 6.25           & 8600        
& 21.5          & 0.42     & 1.26           \nl 
7 & 12 56 40.4                 & 27 54 55.0    & C
& 4.0                  & 6.25           & 5500        
& 21.5          & 0.38     & 1.33           \nl 
8 & 12 54 59.2                & 27 38 09.0     & C
& 4.0                  & 6.25           & 7300        
& 21.5          & 0.40     & 1.31           \nl 
9 & 12 57 31.5                  & 28 18 24.0   & C
& 4.0                  & 6.25           & 9600        
& 21.5          & 0.40     & 1.29          \nl 
10 & 12 55 38.4                  & 28 05 20.0  & C
& 4.0                  & 6.25           & 7300        
& 21.5          & 0.40     & 1.32           \nl 
  &                              &             & D
& 15.0                 & 6.25           & 7300
& 21.5          & 0.20     & 5.46           \nl
11 & 13 01 26.0                 & 28 27 14.0   & C
& 4.0                  & 6.25            & 7000    
& 21.5          & 0.37     & 1.31           \nl 
12 & 12 57 15.4                 & 28 54 12.0   & C
& 4.0                  & 6.25            & 5400      
& 21.5          & 0.40     & 1.28           \nl

\enddata

\end{deluxetable}

\newpage

\begin{deluxetable}{lllllcllrc}
\footnotesize
\tablenum{2}
\tablecaption{Galaxies detected in \hi. \label{tbl-2}}
\tablewidth{0pt}
\tablehead{
  \colhead{(1)}  & \colhead{(2)}        & \colhead{(3)}        
& \colhead{(4)}  & \colhead{(5)}        & \colhead{(6)}
& \colhead{(7)}  & \colhead{(8)}        & \colhead{(9)}  \nl 
  \colhead{CGCG}    & \colhead{Other name} &\colhead{${\alpha}_{1950}$} 
& \colhead{${\delta}_{1950}$}               
& \colhead{Config}  & \colhead{Beam}       & \colhead{Morph}   
& \colhead{B$_{\mathrm T}$}                & \colhead{m$_{\mathrm {UV}}$-b} \nl
  \colhead{}        & \colhead{}           &\colhead{h m s}          
&\colhead{$^{\circ}$ $^{\prime}$ $^{\prime\prime}$}
& \colhead{}        &\colhead{\prin} 
& \colhead{Type}    & \colhead{}           & \colhead{} {}  
}
\startdata
 (1) 160--026 & IC~3913      & 12 54 03.0      & 27 33 40.7 & C          
& 30.7$\times$27.2 & Sc$^{1}$ & 15.4           & -0.46 &  \nl

 (2) 160--055 & NGC~4848     & 12 55 40.7      & 28 30 45.0 & C          
& 30.5$\times$27.5 & Scd      & 14.4           & -0.70 & \nl

 (3) 160--058 &              & 12 55 44.7      & 28 58 40.9 & C        
& 30.5$\times$27.5 & Sc       & 15.5           & 0.27  &  \nl

 (4)         & KUG~1255+275  & 12 55 53.4      & 27 34 49.1 & C     
& 30.7$\times$27.2 & Irr      & 16.3           & -0.92 &  \nl

 (5) 160--067 & Mrk~057      & 12 56 12.0      & 27 26 44.4 & C        
& 30.7$\times$27.2 & Sa$^{1}$ &15.1            & -0.90   &  \nl

 (6) 160--073 & Mrk 058      & 12 56 40.4      & 27 54 48.9 & C          
& 30.3$\times$27.6 & Sb       & 15.2           & -0.30  &   \nl

 (7) 160--076 &              & 12 57 15.7      & 28 54 01.3 & C         
& 29.9$\times$26.9 &Sbc       & 15.7           & -1.35  &  \nl

 (8) 160--086 &              & 12 58 08.9      & 27 54 24.8& C+D         
& 35.2$\times$33.0 & Sb       & 15.6           & -0.94  &  \nl

 (9) 160--252 & IC 4040      & 12 58 13.3      & 28 19 35.0  & C+D         
& 40.7$\times$38.0 & Sdm       & 15.4          & -0.59  &   \nl

(10) 160--088 & IC~842       & 12 58 15.5      & 29 17 18.5 & C+D         
& 36.6$\times$35.4 & Sc$^{1}$ & 14.7           &   &  \nl

(11)         & KUG~1258+287  & 12 58 16.2      & 28 47 20.0 & C         
& 30.2$\times$26.7 &S*      & 16.4             & -1.11  &  \nl

(12) 160--257 & NGC~4907     & 12 58 24.3      & 28 25 38.3 & C+D         
& 39.8$\times$34.5 & Sb      & 14.5            & 0.70  &  \nl

(13) 160--260 & NGC~4911     & 12 58 31.5      & 28 03 34.1 & C+D         
& 40.7$\times$38.0 & Sb      & 13.6            & 0.75  &  \nl

(14)         & FOCA~195     & 12 58 50.7      & 28 47 26.2 & C   
& 30.2$\times$26.7 &        & 17.3$^{1}$      & -1.47  &  \nl

(15) 160-096 & NGC~4922/     & 12 59 $00.2^{1}$& 29 34 $36^{1}$ &C+D    
& 36.6$\times$35.4 & S0$^{1}$& 13.9            &   &  \nl

             & KPG~363--B   & 12 59 $01.4^{1}$& 29 34 $58^{1}$ & C+D  
& 36.6$\times$35.4 & Sb$^{2}$& 14.8            &   &  \nl

(16) 160--098 &             & 12 59 00.8       & 28 56 45.5  & C         
& 30.2$\times$26.7 & Sbc     & 15.3            & -0.13  &  \nl

(17) 160--095 & NGC~4921    & 12 59 01.6       & 28 09 17.2  & C+D       
& 39.8$\times$34.5 & Sb      & 13.3            & 1.35   &  \nl

(18) 160--102 & IC~4088     & 12 59 19.3       & 29 18 47.2  & C+D       
& 36.6$\times$35.4 & Sb$^{1}$& 14.7            &     &  \nl

(19) 160--106 & NGC~4926--A  & 12 59 43.3       & 27 55 00.7  & C+D       
& 35.2$\times$33.0 & Sa      & 15.4            & 0.21   & 
\enddata

\end{deluxetable}

\newpage

\begin{deluxetable}{llrcccccrr}
\footnotesize
\tablenum{3}
\tablecaption{HI parameters \label{tbl-3}}
\tablewidth{0pt}
\tablehead{
  \colhead{(1)}  & \colhead{(2)}        & \colhead{(3)}        
& \colhead{(4)}  & \colhead{(5)}        & \colhead{(6)}
& \colhead{(7)}  & \colhead{(8)}        & \colhead{(9)}   & \colhead{(10)} \nl 
  \colhead{CGCG} & \colhead{Other name} & \colhead{Field} 
& \colhead{v$_{\mathrm {c}}$}           & \colhead{${\Delta}$v} 
& \colhead{H{\sc i} flux}               & \colhead{F$_{\mathrm {cont}}$}
& \colhead{M$_{\mathrm {HI}}$}          & \colhead{\defhi}          
& \colhead{r/r$_{\mathrm A}$}       					   \nl
  \colhead{}                            & \colhead{}           
& \colhead{}      			& \colhead{km~s$^{-1}$}       
& \colhead{km~s$^{-1}$} 		& \colhead{Jy~km~s$^{-1}$}  
&\colhead{mJy}                          & \colhead{$10^{9}$M$_{\odot}$}
& \colhead{}                            &\colhead{}  			   \nl
}
\startdata
 (1) 160--026 &IC~3913 &8& 7538 & 260 & 1.08$\pm$0.06 & 1.73 & 1.25 &0.04 
& 0.80\nl

 (2) 160--055 &NGC~4848&3& 7041 & 130 & 0.37$\pm$0.03 &16.56 & 0.43 &1.14 
& 0.36\nl

 (3) 160--058 &        &3& 7614 & 281 & 1.67$\pm$0.06 & 4.71 & 1.93 &0.12 
& 0.68\nl

 (4)           &KUG~1255+275&8& 7398 & 281 & 0.65$\pm$0.05 & 2.02 &0.75 &-0.17
&0.59\nl

 (5) 160--067 &Mrk~057 &8& 7657 & 194 & 1.07$\pm$0.04 & 3.34 & 1.24 &-0.44
& 0.68\nl

 (6) 160--073 &Mrk~058 &7& 5425 & 64  & 0.17$\pm$0.02 & 5.46 & 0.20 & 0.51
& 0.28\nl

 (7) 160--076 &        &12&5390 & 235 & 2.00$\pm$0.06 & 5.20 & 2.31 &-0.65
& 0.55\nl

 (8) 160--086 &        &4& 7481 & 130 & 0.15$\pm$0.02 & 3.76 & 0.17 & 0.58
& 0.33\nl

 (9) 160--252 &IC~4040 &1& 7758 & 347 & 0.29$\pm$0.06 & 15.0 & 0.33 & 0.61
& 0.20\nl

(10) 160--088 &IC~842  &5& 7281 & 389 & 1.50$\pm$0.04 & 2.67 & 1.73 & 0.31
& 0.90\nl

(11)           &KUG~1258+287&6& 8884 & 306 & 1.17$\pm$0.06 & 7.12 & 1.35&0.01
&0.51\nl

(12) 160--257 &NGC~4907&2& 5821 & 214 & 0.21$\pm$0.03 &     & 0.24 & 0.99
& 0.27\nl

(13) 160--260 &NGC~4911&1& 7997 & 348 & 0.80$\pm$0.07 & 11.92& 0.93 & 0.58
&0.28\nl

(14)          &FOCA~195    &6& 8426 & 218 & 0.57$\pm$0.04 &      & 0.66 &-0.11
&0.55\nl

(15) 160--096 &NGC~4922/&5& 6892 & 130 & 0.12$\pm$0.02 &27.47 & 0.14 & 1.28
& 1.17\nl

              &KPG~363--B&5&7086 &     &      &27.47 &       &     
& 1.17\nl

(16) 160--098 &        &6&8764 & 153 & 0.63$\pm$0.04 & 5.67 & 0.73  & 0.16
& 0.69\nl

(17) 160--095 &NGC~4921&2& 5479 & 214 & 0.61$\pm$0.03 & 4.87 & 0.70 & 1.11
& 0.34\nl

(18) 160--102 &IC~4088 &5& 7108 & 475 & 3.46$\pm$0.07 & 8.50 & 4.00 &-0.01
& 0.98\nl

(19) 160--106 &NGC~4926--A &4& 6876 & 216 & 0.52$\pm$0.03 & 3.09 & 0.60 &0.16
&0.52\nl 
\enddata

\end{deluxetable}

\newpage

\begin{deluxetable}{llrcrlccccl}
\footnotesize
\tablenum{4}
\tablecaption{Galaxies non-detected in HI \label{tbl-4}}
\tablewidth{0pt}
\tablehead{
  \colhead{(1)}  & \colhead{(2)}      & \colhead{(3)}        
& \colhead{(4)}  & \colhead{(5)}      & \colhead{(6)}
& \colhead{(7)}  & \colhead{(8)}      & \colhead{(9)}     & \colhead{(10)} \nl 
  \colhead{Name} & \colhead{Other}     
& \colhead{${\alpha}_{1950}$}         & \colhead{${\delta}_{1950}$}
& \colhead{field} 
& \colhead{Morph}  		      & \colhead{v$_{\mathrm {hel}}$}   
& \colhead{rms}   		      & \colhead{M$_{\mathrm {HI}}$} 
& \colhead{\defhi}  							   \nl
  \colhead{}     		      & \colhead{name}  
& \colhead{h m s}           	      & \colhead{\degree~\prim~\prin}
& \colhead{} 			      & \colhead{type}  
& \colhead{\km}                       & \colhead{mJy/beam} 
& \colhead{10$^{8}$\msolar}           & {}       
}
\startdata
160--031 & MCG~5-31-19 & 12 54 24.1 & 27 21 50 & 8 & Sa$^{2}$ & 6849    
&0.80    & $<$1.2      & $>$1.07      \nl
160--043 & UGC~8071    & 12 55 04.0 & 28 27 40 & 3 & Sa       & 7069       
& 0.78   & $<$1.2      & $>$1.22      \nl
160--061 & NGC~4851    & 12 55 56.8 & 28 25 06 &10 & SBa      & 7781 
& 0.35   & $<$0.5      & $>$1.14      \nl
160--064 & Mrk~056     & 12 56 09.4 & 27 32 10 & 8 & Sa*      & 7392       
& 0.76   & $<$1.1      & $>$0.69      \nl
160--069 & IC~3943     & 12 56 11.5 & 28 22 59 &10 & S0/a     & 6704       
& 0.31   & $<$0.5      & $>$1.32      \nl
160--213 & NGC~4858    & 12 56 36.9 & 28 23 14 & 9 & Sc (SB)  & 9456       
& 0.62   & $<$0.9      & $>$0.93      \nl
160--216 & IC 3955     & 12 56 41.0 & 28 15 59 &10 & SB0/a    & 7650
& 0.22   & $<$0.3      & $>$1.01      \nl
         & KG1258+277  & 12 58 04.0 & 27 47 06 & 4 & S0*      & 7651       
& 0.62   & $<$0.9      & $>$0.70      \nl
160--261 &             & 12 58 35.0 & 28 10 13 & 4 & S0/a     & 6866       
& 0.40   & $<$0.6      & $>$1.27      \nl
160--108 & MCG~5-31-108& 12 59 48.1 & 28 29 07 & 1 & Sb       & 8309       
& 0.80   & $<$1.2      & $>$0.73      \nl
160--112 & IC~4106     & 13 00 14.6 & 28 23 05 &11 & S0/a     & 7443       
& 0.68   & $<$1.0      & $>$1.14      \nl
160--124 & NGC~4944    & 13 01 26.0 & 28 27 14 &11 &S0/a$^{3}$& 7034       
& 0.37   & $<$0.6      & $>$1.90      \nl \hline

\enddata

\end{deluxetable}

\newpage

\begin{deluxetable}{llrcrlcccc}
\footnotesize
\tablenum{5}
\tablecaption{SB and PSB galaxies non-detected in HI \label{tbl-5}}
\tablewidth{0pt}
\tablehead{
  \colhead{(1)}  & \colhead{(2)}      & \colhead{(3)}        
& \colhead{(4)}  & \colhead{(5)}      & \colhead{(6)}
& \colhead{(7)}  & \colhead{(8)}      & \colhead{(9)}     \nl 
  \colhead{ID} & \colhead{Other}      & \colhead{field} 
& \colhead{${\alpha}_{1950}$}         & \colhead{${\delta}_{1950}$} 
& \colhead{Morph}  		      & \colhead{v$_{\mathrm {hel}}$}   
& \colhead{rms}   		      & \colhead{M$_{\mathrm {HI}}$}   \nl
  \colhead{}     		      & \colhead{name}  
& \colhead{}      		      & \colhead{h m s}           
& \colhead{\degree~\prim~\prin}	      & \colhead{type}  
& \colhead{\km}                       & \colhead{mJy/beam} 
& \colhead{10$^{8}$\msolar}           & {}   
}
\startdata
D 77     & Leda 83676  &10 & 12 54 39.2 & 28 02 35   & S0/a (SB)& 7544
& 0.39   & $<$0.6      &  \nl
D 94     & Leda 83682  &10 & 12 54 52.6 & 28 04 51   & SA0 (PSB)& 7084 
& 0.35   & $<$0.5      &  \nl
D 112    & Leda 83684  &10 & 12 54 56.4 & 28 09 02   & SB0 (PSB)& 7428  
& 0.27   & $<$0.4      & \nl     
D 21 	 & MCG 5-31-037& 8 & 12 55 36.3 & 27 45 33   & SBa (PSB) & 7684
& 0.48	 & $<$0.7      &  \nl
D 73 	 & RB 183      & 7 & 12 55 53.9 & 28 01 54   & SA0 (PSB) & 5434
& 0.61	 & $<$0.9      &  \nl
D 44     &KUG1256+278A  &10 & 12 56 03.1 & 27 49 45   & S0  (SB) & 7554 
& 0.37   & $<$0.6      &  \nl
D 43     & NGC~4853    &10 & 12 56 10.1 & 27 51 58   & SA0 (SB) & 7660 
& 0.32   & $<$0.5      &  \nl
D 89     & IC 3949     &10 & 12 56 31.4 & 28 06 12.1 & SA0 (PSB)& 7378
& 0.18   & $<$0.3      & \nl       
D 127    & RB 042      &10 & 12 57 15.4 & 28 14 16   & S0 (PSB) & 7514 
& 0.41   & $<$0.6      &  \nl
D 216    & RB 160      & 1 & 12 57 38.2 & 28 30 34   & Sa  (PSB)& 7684 
& 0.46   & $<$0.7      &  \nl
D 99     & Mrk 060     & 9 & 12 57 45.4 & 28 07 59   & SB0 (PSB) & 9902
& 0.48   & $<$0.7      &  \nl
D 146    & RB 110      & 1 & 12 58 14.1 & 28 17 00   & S0  (PSB) & 7537 
& 0.20   & $<$0.3      &  \nl 
D 61 	 & CGCG 160-104& 4 & 12 59 35.6 & 28 03 04   & SA0 (PSB) & 7102
& 0.44	 & $<$0.7      &  \nl
D 189	 & Leda 83763  & 2 & 12 59 49.9 & 28 22 13   & S0  (PSB) & 5937
& 0.97	 & $<$1.5      &  \nl

\enddata

\end{deluxetable}

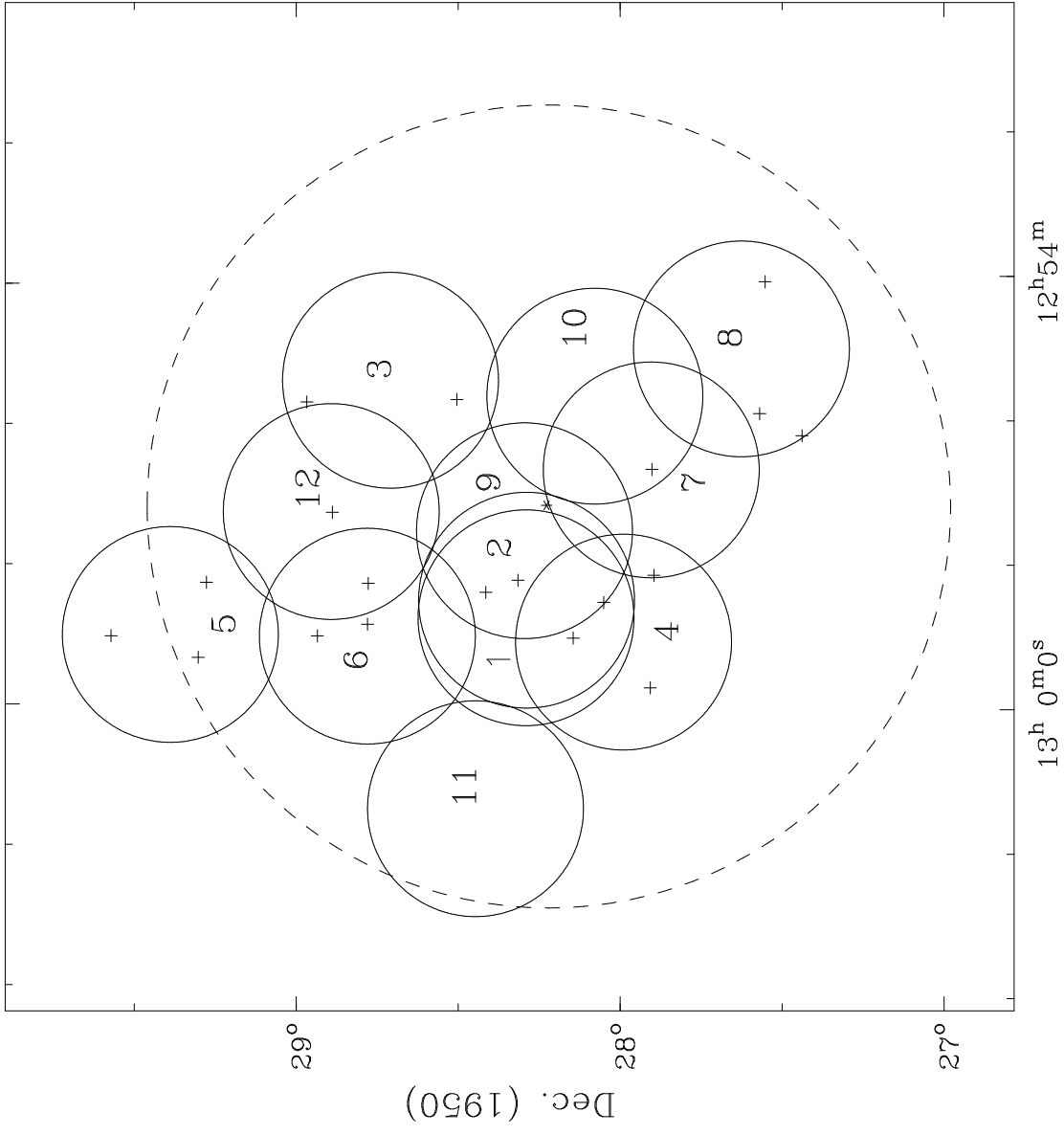
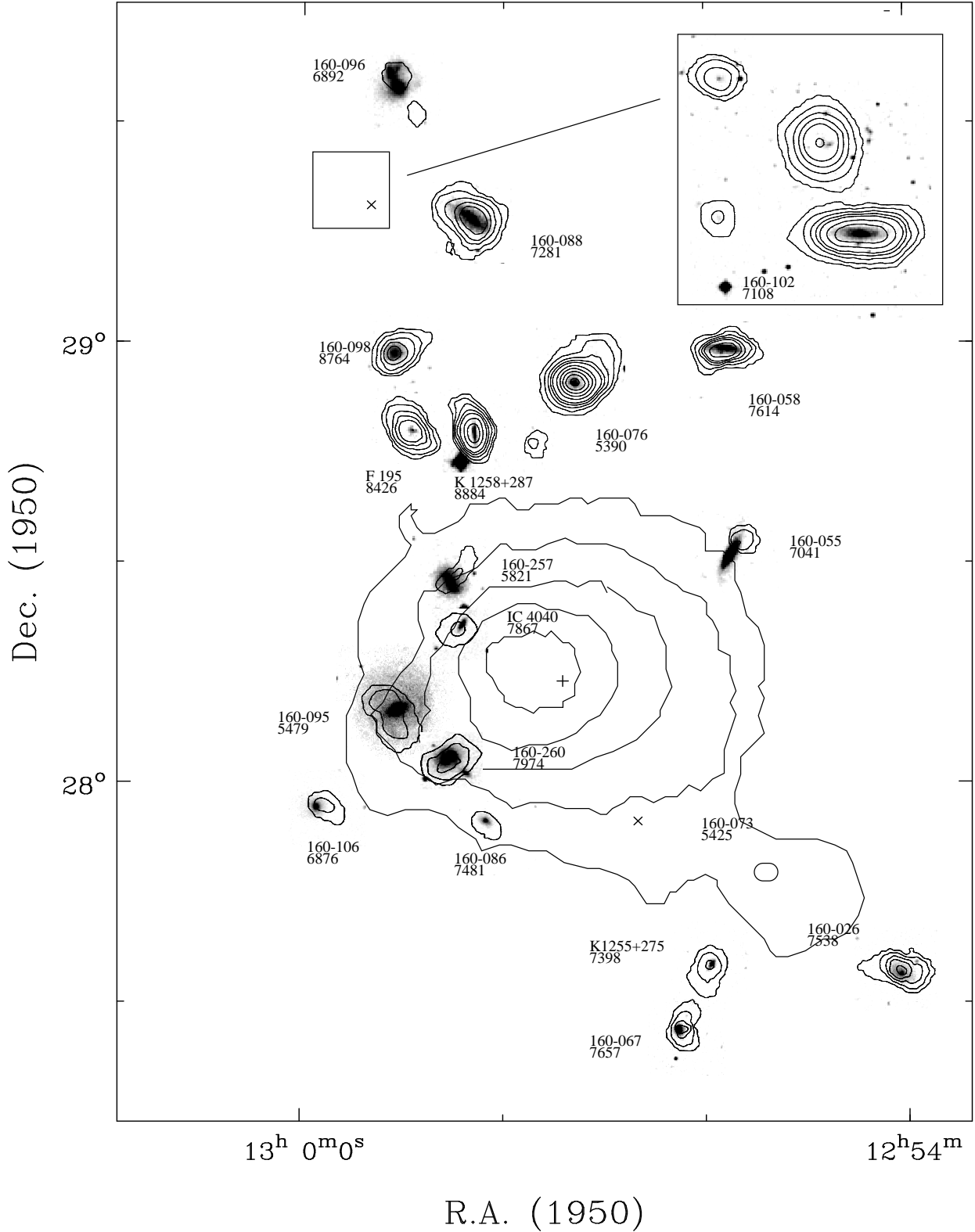
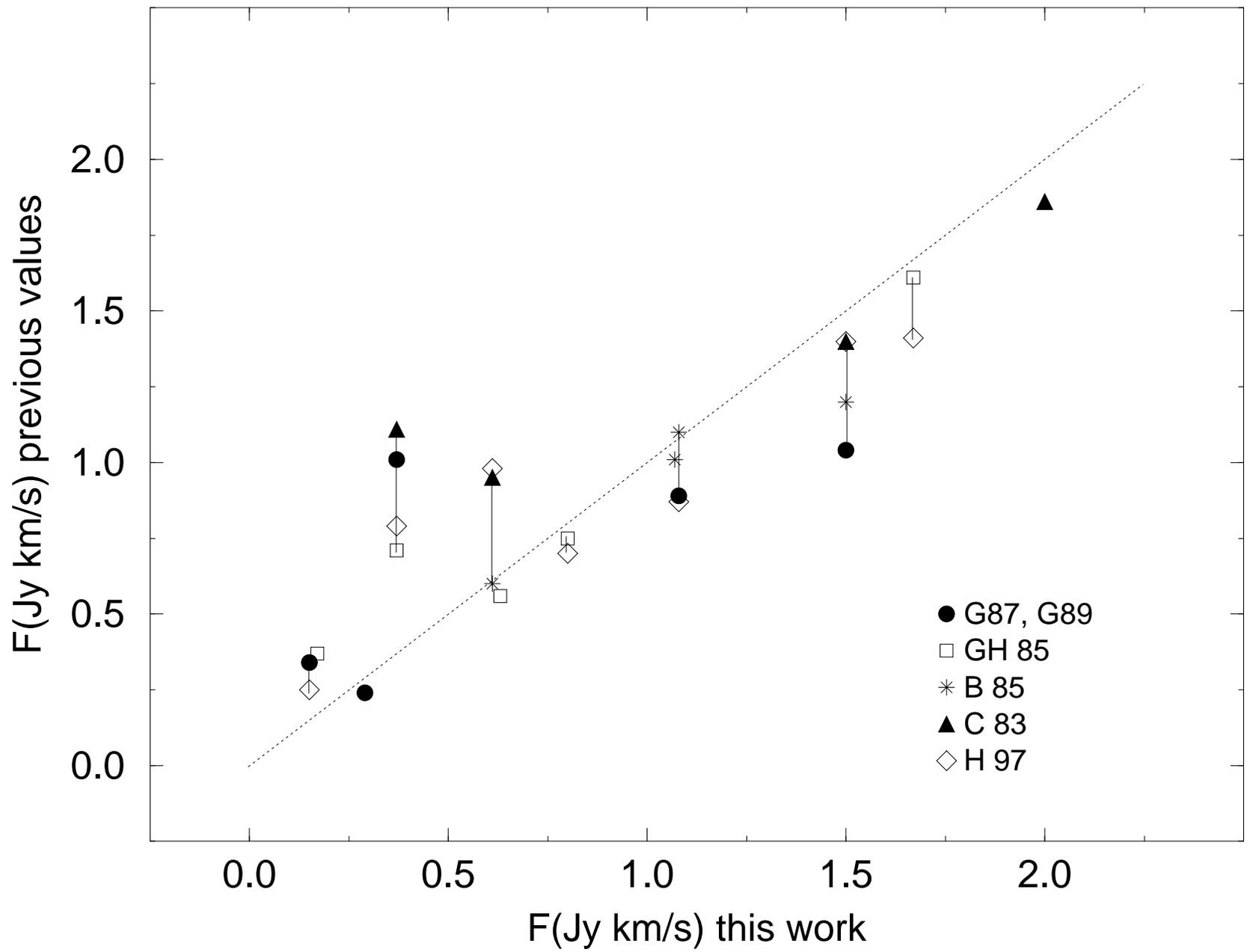
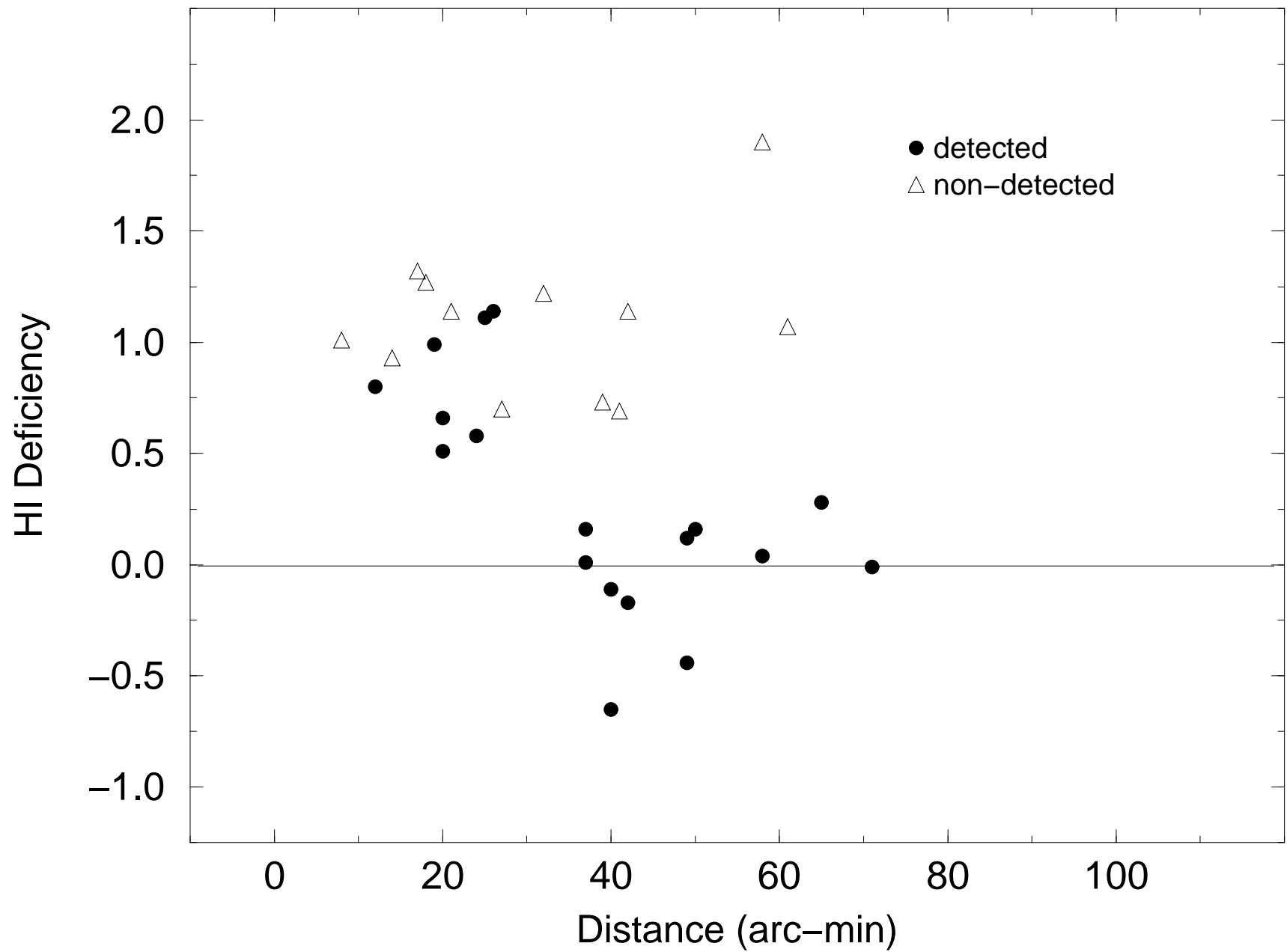
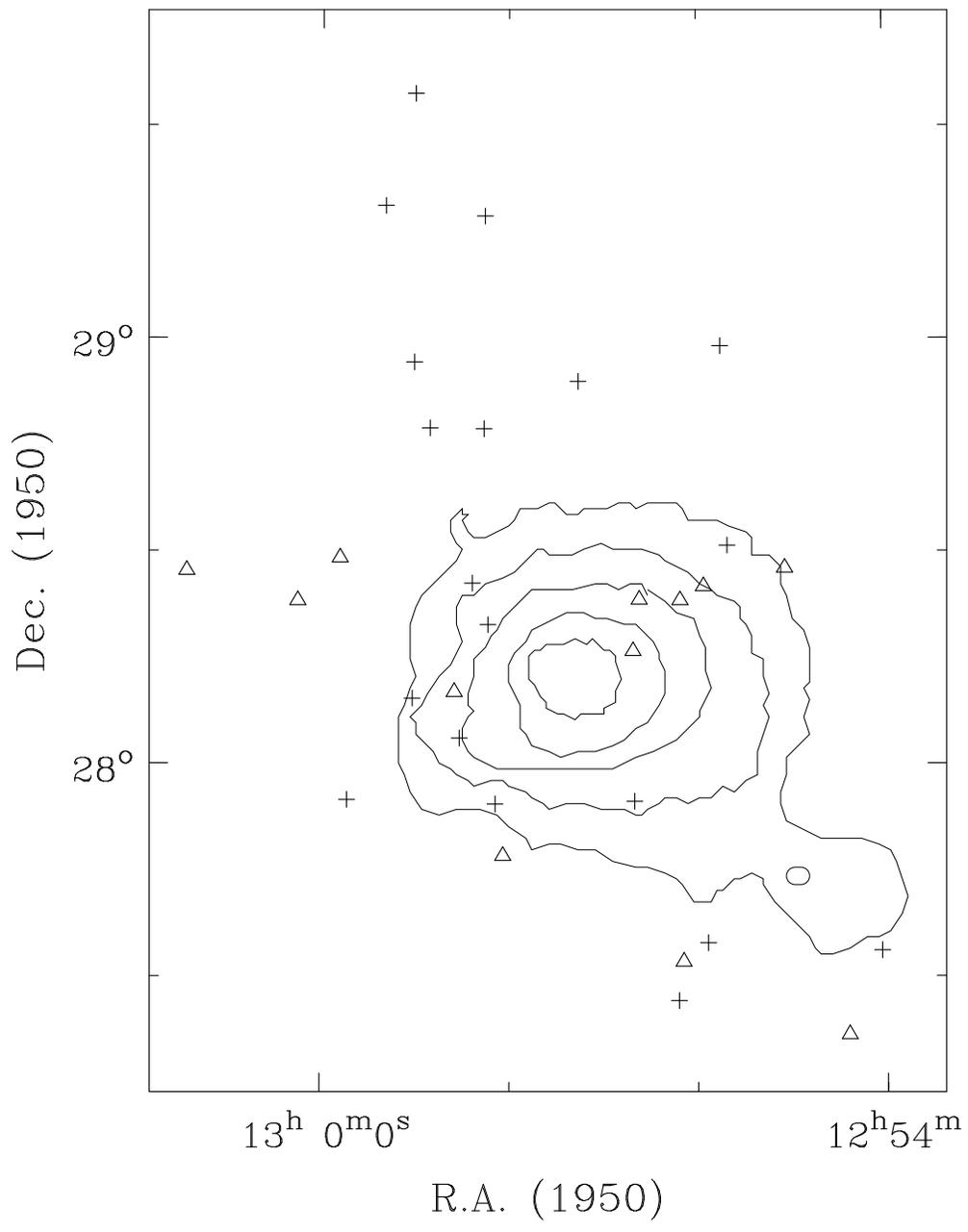
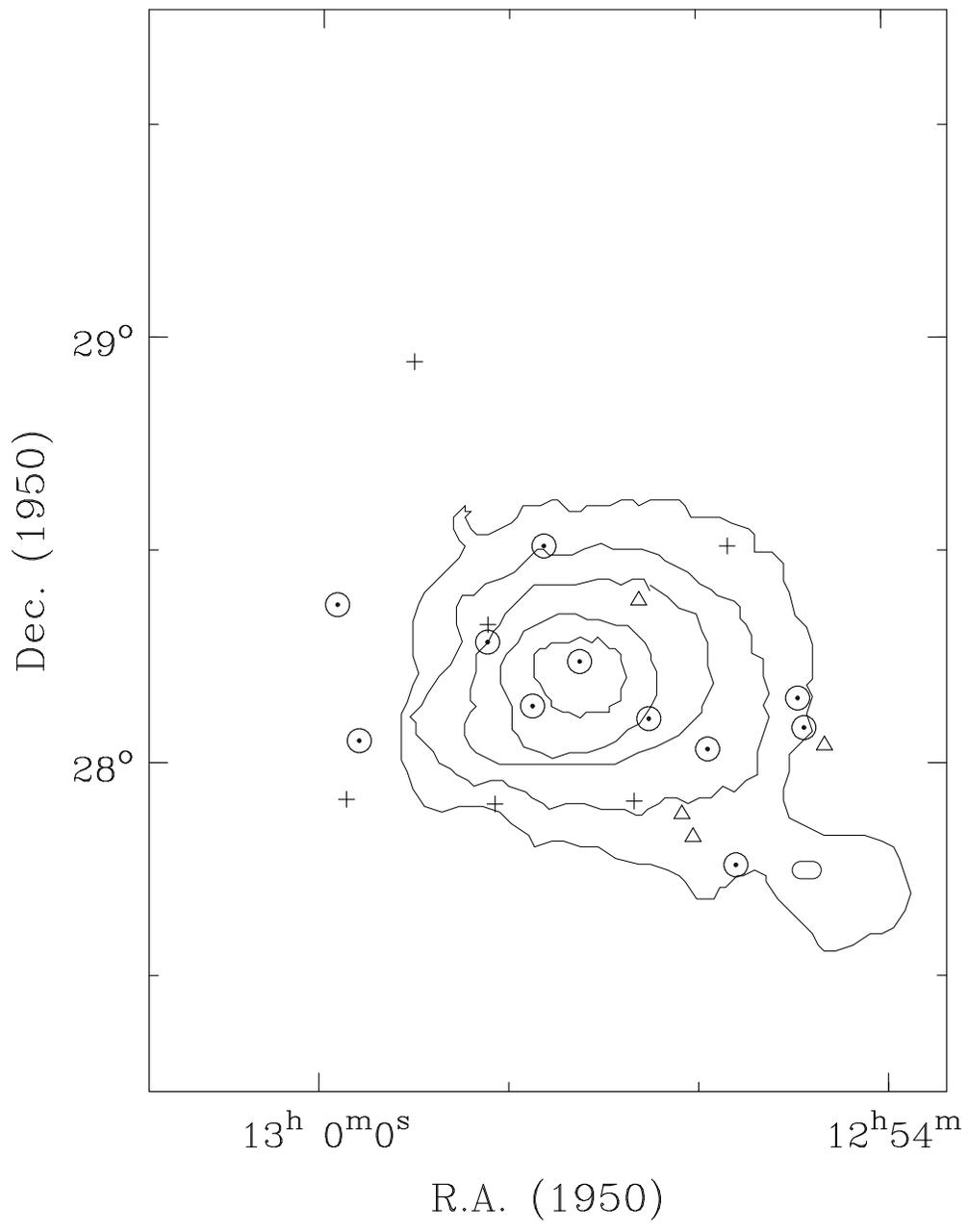
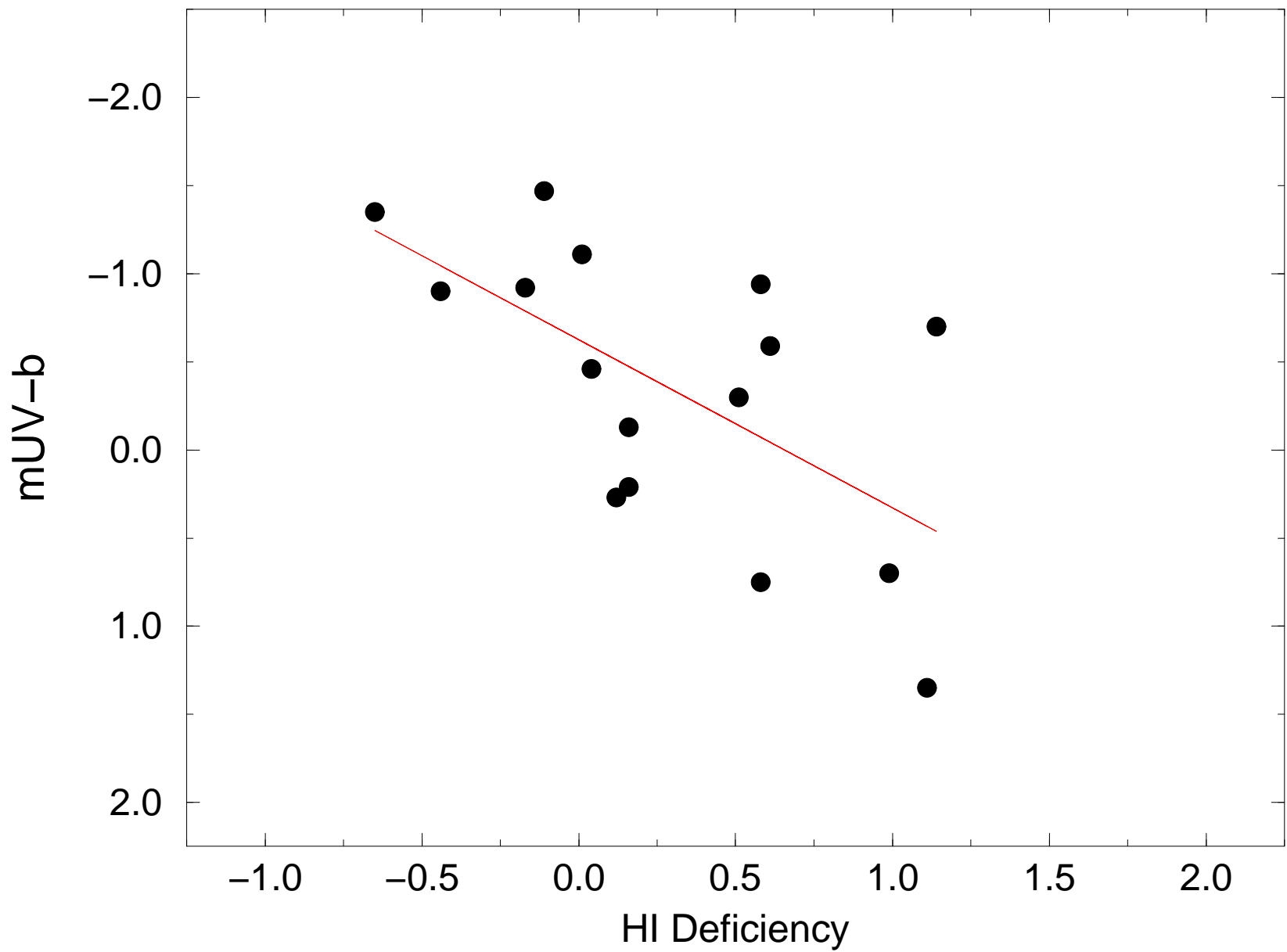
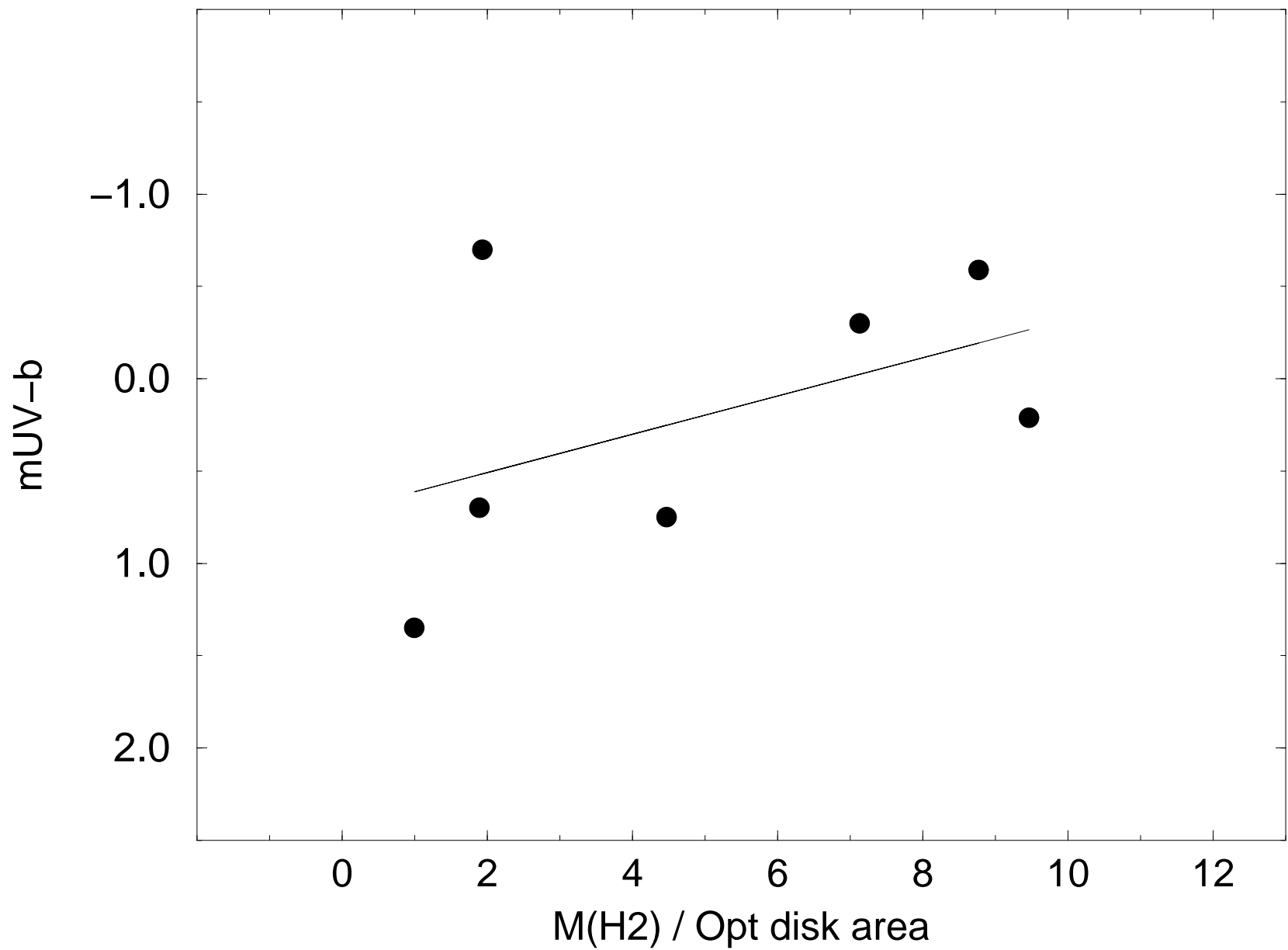
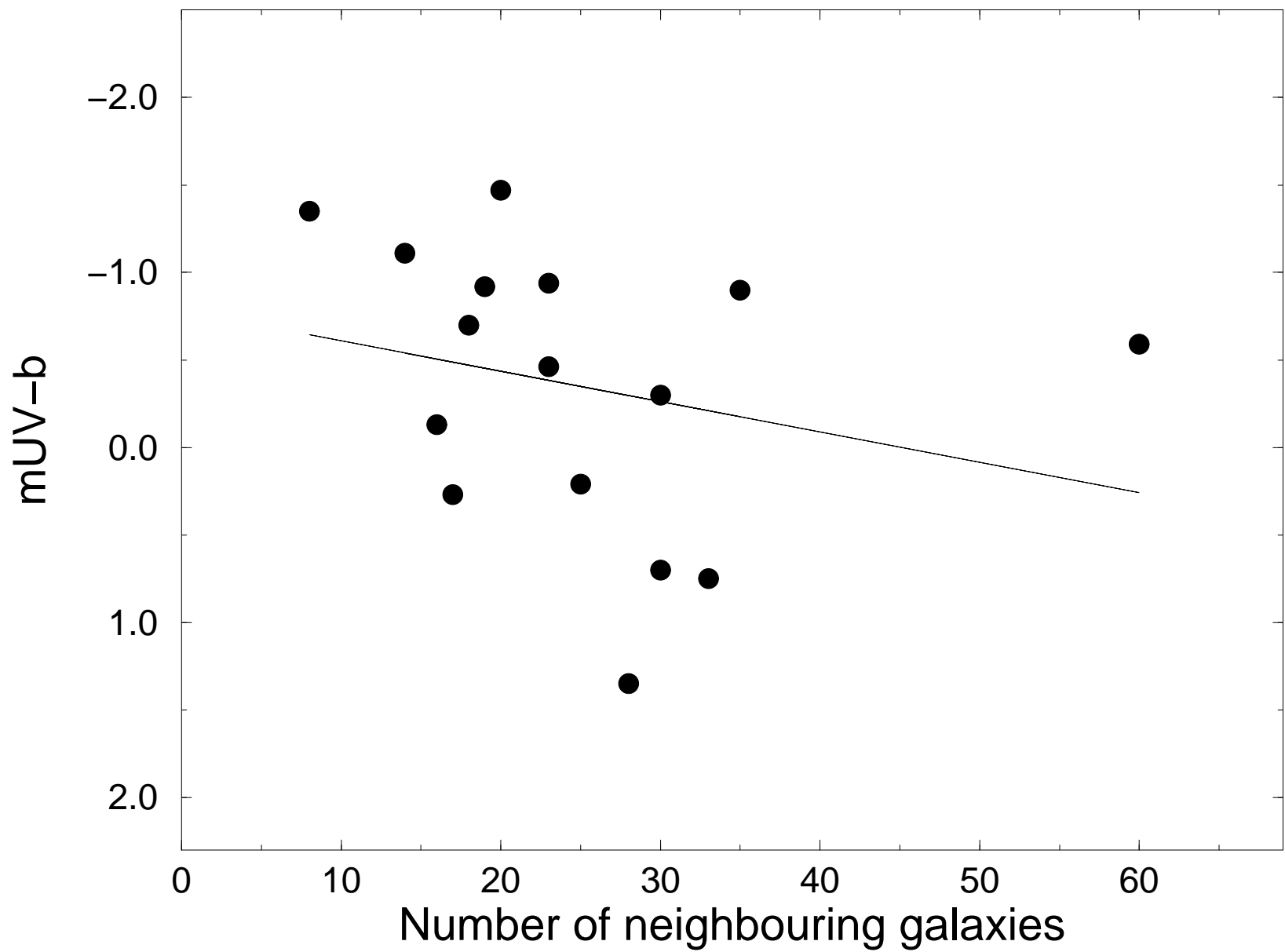
\end{document}